\documentclass[10pt,journal,compsoc]{IEEEtran}

\usepackage{todonotes}
\usepackage{ctable} 

\usepackage{placeins}

\ifCLASSOPTIONcompsoc
  \usepackage{cite} 
\else
  \usepackage{cite}
\fi
\ifCLASSINFOpdf
\else
\fi
\usepackage{amsmath}
\usepackage{amssymb}
\usepackage{algorithm}
\usepackage{algorithmic}
  \usepackage[caption=false,font=footnotesize]{subfig}
\usepackage{fixltx2e}
\usepackage{url}

\newcommand{\eg}{\textit{e.g.}~}
\newcommand{\ie}{\textit{i.e.}~}

\newcommand{\R}{\mathbb{R}}

\DeclareMathOperator*{\argmin}{\textnormal{argmin}}

\newcommand{\brac}[1]{\left(#1\right)}

\newcommand{\abs}[1]{\left| #1 \right|}
\newcommand{\norm}[1]{\left\| #1 \right\|}

\newcommand{\alref}[1]{{\bf Algorithm \ref{#1}}}
\newcommand{\figref}[1]{{\bf Figure \ref{#1}}}
\newcommand{\tabref}[1]{{\bf Table \ref{#1}}}

\newcommand{\appref}[1]{{\bf Appendix \ref{#1}}}
\newcommand{\secref}[1]{{\bf Section \ref{#1}}}

\newcommand{\FDK}{\text{FDK}}
\newcommand{\q}{\mathbf{q}}
\newcommand{\h}{\mathbf{h}}
\newcommand{\y}{\mathbf{y}}
\newcommand{\Fy}{F_{\mathbf{y}}}
\newcommand{\pderiv}[2]{\frac{\partial #1}{\partial #2}}

\usepackage{tikz}
\usetikzlibrary{shapes,arrows,shadows, fit}
\usepackage[switch]{lineno}

\date{\today}
\begin{document}

\title{A computationally efficient reconstruction algorithm for
  circular cone-beam computed tomography using shallow neural networks.}

\author{Marinus~J.~Lagerwerf, Dani\"el~M.~Pelt, Willem~Jan~Palenstijn,
  and~K.~Joost~Batenburg.
  \IEEEcompsocitemizethanks{\IEEEcompsocthanksitem M..J. Lagerwerf,
    D.M. Pelt, W.J. Palenstijn and K.J. Batenburg are with the Computational Imaging group,
    Centrum voor Wiskunde en Informatica, Amsterdam, Science Park 123, 1098
    XG.}%
}

%
%

\markboth{IEEE Transactions on Computational Imaging}%
{Lagerwerf \MakeLowercase{\textit{et al.}}: Fast and memory efficient CCB CT
  reconstruction algorithm using neural networks}

\maketitle


%
\begin{abstract}
Circular cone-beam (CCB) Computed Tomography (CT) has become an integral part of industrial quality control, materials science and medical imaging. The need to acquire and process each scan in a short time naturally leads to trade-offs between speed and reconstruction quality, creating a need for fast reconstruction algorithms capable of creating accurate reconstructions from limited data.

In this paper we introduce the Neural Network Feldkamp-Davis-Kress (NN-FDK) algorithm. This algorithm adds a machine learning component to the FDK algorithm to improve its reconstruction accuracy while maintaining its computational efficiency. Moreover, the NN-FDK algorithm is designed such that it has low training data requirements and is fast to train. This ensures that the proposed algorithm can be used to improve image quality in high throughput CT scanning settings, where FDK is currently used to keep pace with the acquisition speed using readily available computational resources. 

We compare the NN-FDK algorithm to two standard CT reconstruction algorithms and to two popular deep neural networks trained to remove reconstruction artifacts from the 2D slices of an FDK reconstruction. We show that the NN-FDK reconstruction algorithm is substantially faster in computing a reconstruction than all the tested alternative methods except for the standard FDK algorithm and we show it can compute accurate CCB CT reconstructions in cases of high noise, a low number of projection angles or large cone angles. Moreover, we show that the training time of an NN-FDK network is orders of magnitude lower than the considered deep neural networks, with only a slight reduction in reconstruction accuracy.

\end{abstract}


\section{Introduction}\label{sec:intro}
Circular cone-beam (CCB) Computed Tomography (CT) has become an integral part of
non-destructive imaging in a broad spectrum of applications, such as industrial
quality control \cite{giudiceandrea2011high}, materials sciences
\cite{dierick2014recent,bultreys2016fast} and medical imaging
\cite{ford2002cone,galicia2017clinical}. Limitations on the scanning process
caused by the need to scan a large number of objects in a short amount of time
lead to measurements with a low number of projection angles or high noise
levels. Additionally, CT reconstruction has become a \emph{big data} problem due
to the development of readily available high-resolution CT-scanners
\cite{xre2019unitom, xre2019dynatom, canon2019precision}. This stresses the need for
computationally efficient reconstruction methods that are applicable to a broad
spectrum of high-resolution problems and produce accurate results from data with
a high noise levels, low number of projection angles or large cone angles.

In practice, if computational efficiency is a constraint and especially for high-resolution problems, \emph{direct methods}  (\eg the filtered backprojection (FBP) algorithm \cite{natterer2001mathematics}, the Feldkamp-Davis-Kress (FDK) algorithm \cite{feldkamp1984practical} and the Katsevich algorithm \cite{katsevich2003general}) are still the common choice of reconstruction method \cite{pan2009commercial}. While \emph{iterative methods} have been shown to be more accurate for noisy and limited data problems \cite{rudin1992nonlinear, bredies2010total, sidky2008image, jia2010gpu, 0031-9155-59-12-2997, elbakri2003efficient}, they have a significantly higher computational cost. Consequently there have been efforts to improve the accuracy of direct methods by computing data-specific or scanner-specific filters \cite{l2012filtered, nielsen2012filter, batenburg2012fast, pelt2014improving, lagerwerf2020automated}. Although these strategies do improve the reconstruction accuracy, they also add significant computational effort or are specific to one modality, \eg tomosynthesis \cite{kunze2007filter}.

An emerging approach for improving direct methods is to use \emph{machine learning} to remove artifacts from the reconstructions. The idea is to use
high-quality reconstructions to train a neural network that removes artifacts
from low-quality reconstructions using a supervised learning approach. This
\emph{post-processing} approach has shown promising results for computed
tomography using deep neural networks (DNNs) \cite{jin2017deep, pelt2018improving, kida2018cone}. Deep neural network structures contain a
large number of layers, leading to millions of trainable parameters and
therefore require a large amount of training data \cite{pelt2018mixed}. This is
problematic in CT imaging, since there is often a limited amount of training data
available, \eg due to scanning time, dose, and business-related concerns. Moreover, for
the available data there are often no reference datasets or annotations
available \cite{wang2018image}. The large amount of training data and large
number of parameters also lead to long training times. While for standard 2D
networks the training time ranges between a couple of hours and a couple of days
(see \secref{sec:train_time}), for 3D networks the training time becomes prohibitively long \cite{cciccek20163d} (\ie weeks). Therefore, to
apply post-processing to 3D problems the reconstruction volume can be considered
as a stack of 2D problems \cite{ronneberger2015u, pelt2018improving} for which
one 2D network is trained and then applied in a \emph{slice-by-slice} fashion to
the 3D volume. Although this strategy reduces the training time and the training data constraints,
applying a 2D network to all slices can still be computationally intensive due
to the number of slices in the 3D volume. A more in-depth discussion on current developments related to machine learning methods in CT imaging is given in \secref{sec:rel_work}.

In this work we propose the Neural Network FDK (NN-FDK) reconstruction
algorithm. It is a direct reconstruction method that is designed to produce accurate results from noisy data, data with a low number of projection angles, or a large cone angle, but still maintains a similar computational efficiency and scalability as the standard FDK algorithm. Moreover, the algorithm has a fast training procedure, and requires a limited amount of training data.

The NN-FDK algorithm is an adaptation of the
standard FDK algorithm using a shallow multilayer
perceptron network \cite{bishop2006pattern} with one fully connected hidden layer,
a low number of trainable parameters and low memory constraints.
We will show it is possible to interpret the weights of the first layer
of the perceptron network as a set of learned filters for the FDK
algorithm. We can then use the FDK algorithm to evaluate the network
efficiently for all voxels simultaneously to arrive at an accurate
reconstruction for the CCB CT problem.


The NN-FDK algorithm is an extension of the method proposed in
\cite{pelt2013fast} for the Filtered Backprojection (FBP) algorithm
\cite{natterer2001mathematics}. The derivation of the approach outlined in
\cite{pelt2013fast} relies on the shift-invariance property of the FBP
algorithm. We will show that, although the FDK algorithm does not have this
shift-invariance property, we can derive a similar method for the FDK algorithm.
Moreover, the proposed strategy can be extended to any linear filtered
backprojection type reconstruction method.

Using both simulated and experimental data,
we compare the proposed method with the standard FDK algorithm, SIRT \cite{van1990sirt}
with a nonnegativity constraint (SIRT$^+$), which is a commonly used iterative
algorithm for CT problems, and two 2D deep neural networks (U-net \cite{ronneberger2015u} and MSD
\cite{pelt2018improving}) trained
to remove reconstruction artifacts from slices of standard FDK reconstruction. We show
that the NN-FDK algorithm is faster to evaluate than all but the standard FDK
algorithm and orders of magnitude faster to train than the considered DNNs, with
only a slight reduction in reconstruction accuracy compared to the DNNs.

The paper is structured as follows. In \secref{sec:method} we give definitions
and introduce our method. In \secref{sec:experiments} we introduce the data and
the parameters used for the experiments. The experiments and their results are shown
and discussed in \secref{sec:results}. The paper is summarized and concluded in
\secref{sec:conclusions}.

\section{Related work}\label{sec:rel_work}
Using \emph{machine learning} methods is an emerging approach in CT imaging
\cite{wang2018image}. \emph{Deep learning} methods have shown
promising results for many applications within the development of CT reconstruction methods \cite{kang-2017-deep-convol}. For the sake of exposition, we split these machine learning approaches into two categories: (i) Improving standard reconstruction methods by replacing components of the reconstruction method with networks specifically trained for the application; and (ii) 
improving the image quality of reconstructions computed with existing reconstruction methods by training neural networks to perform \emph{post-processing} in order to remove artifacts or reduce noise. 

Examples of the first strategy (improving standard reconstruction methods) applied to iterative methods are the learned primal-dual reconstruction algorithm \cite{adler2017solving, adler2018learned}, variational networks \cite{kobler2017variational, hammernik2018learning},  plug and play priors \cite{venkatakrishnan-2013-plug-and, romano-2016-littl-engin, reehorst-2018-regul-by-denois}, and learned regularizers \cite{lunz2018adversarial, mukherjee2020learned}. These methods achieve promising results in reconstruction accuracy and generalizability. However, their high computational cost limits the applicability if high throughput is required. 
Examples for this strategy applied to direct methods are the NN-FBP method \cite{pelt2013fast}, and also the NN-FDK method introduced in this paper. These methods are designed to improve the image quality of direct methods for data with limitations (\eg data with noise or a low number of projection angles) while maintaining their computational efficiency. 

Examples of the second strategy (learned post-processing) have demonstrated substantial improvements in reconstruction quality for CT imaging  \cite{ronneberger2015u, kang-2017-deep-convol, pelt2018mixed, jin2017deep}.  This is aided by the fact that the post-processing problem can be viewed as a classic imaging problem -- \eg denoising, segmentation, inpainting, classification -- for which many effective machine learning methods have already been developed \cite{shelhamer-2017-fully-convol, perone-2018-spinal-cord, zhang-2017-beyon-gauss-denois}. Although the general trend is towards deeper networks to make such networks more expressive \cite{ye2018deep}, this can lead to problems with scalability for large 3D image datasets.

The rise in popularity of machine learning in CT is driven by the increased computational possibilities and although these advances are sufficient to handle most 2D problems, scaling towards 3D problems can be problematic, due to memory constraints. This is illustrated in \figref{fig:mem_con} in \secref{sec:mem_con}, where we plotted the memory constraints for applying a 2D and 3D U-net and MSD network in terms of gigabytes (GiB) of memory as a function of the size of the image. This shows that in theory one could apply a 2D MSD network to images of $7500\times7500$ pixels (with a 24GiB GPU), but in 3D this limit lies around $400\times400\times400$ voxels. Considering that CT problems range between $256\times256\times256$ (small image size) up to $4096\times 4096\times 4096$ images, this gives an indication that scalability can become an issue, especially for 3D problems. 

When applying machine learning techniques for improving the reconstruction quality in CT, a balance must be struck between image quality, running time, and memory requirements. Here we propose a method that achieves relatively high accuracy, while also being computationally efficient and scalable.

\section{Method}\label{sec:method}
The NN-FDK algorithm is a reconstruction algorithm with a machine learning
component, meaning that a number of parameters of the reconstruction algorithm
are optimized through \emph{supervised learning} \cite{anthony2009neural}.
Similar to the network presented in \cite{pelt2013fast}, the \emph{NN-FDK
  network} is a two layer neural network with a hidden layer and an output
layer. We design the network such that it
reconstructs one single voxel, but handles all voxels in a similar manner. This means
that we only have to train one network for a full reconstruction. We consider
the NN-FDK algorithm to have three parts: The \emph{NN-FDK network}, the
\emph{NN-FDK reconstruction algorithm} and the \emph{training process}.

We introduce the reconstruction problem, FDK algorithm, a filter approximation
method and the definition of a perceptron in \secref{sec:prelim}. In
\secref{sec:deriv} we give the \emph{NN-FDK reconstruction algorithm} and derive
from this algorithm the \emph{NN-FDK network}. The input of the network that is
needed in the \emph{training process} is a pre-processed version of the input of
the reconstruction algorithm. In \secref{sec:training}, we discuss how to
compute this pre-processing step for all voxels simultaneously and we introduce
the optimization problem and related notation for the training process. Lastly,
we summarize and discuss the characteristics of the method in \secref{sec:theo_comp}.

\subsection{Preliminaries}\label{sec:prelim}
\subsubsection{Reconstruction problem} In this paper we focus exclusively on the
circular cone-beam (CCB) geometry, where the object rotates with respect to a point
source and a planar detector, acquiring 2D cone-beam projections. The
reconstruction problem for the CCB geometry can be modeled by a system of linear
equations
\begin{align}
  W\mathbf{x}=\mathbf{y},\label{eq:IP}
\end{align}
where $\mathbf{x}\in\mathbb{R}^n$ is the vector describing the reconstruction (\ie every element coincides with a voxel value), $\mathbf{y}\in\mathbb{R}^m$ is
the vector describing the measured projection data, and
${W}\in\mathbb{R}^{m\times n}$ is a discretized version of the \emph{cone-beam
  transform} or \emph{forward projection}. For the sake of simplicity we assume
that the volume consists of $n = N{\times}N{\times}N$ voxels and the detector
consists of $N{\times}N$ pixels. We denote the number of angles with $N_a$, so
we have $m=N_a\times N\times N$.

\subsubsection{FDK algorithm \& filter approximation}\label{sec:FDK}
The FDK algorithm, as presented in \cite{feldkamp1984practical}, is a filtered
backprojection-type algorithm that solves the CCB reconstruction problem
\eqref{eq:IP} approximately. First, for each projection angle, it applies a
\emph{reweighting} step, $r:\R^{N_a\times N\times N}\rightarrow\R^{N_a\times
  N\times N}$, that adapts the cone-beam data such that it approximately behaves
as fan-beam data. Second, it applies a \emph{filtering} step, that convolves the
data with a one-dimensional \emph{filter} $\mathbf{h}$ in a line-by-line
fashion, $(- * -)_{\text{1D}}:\R^{2N}\times \R^{N_a\times N\times
  N}\rightarrow\R^{N_a\times N\times N}$. Last, it applies a
\emph{backprojection} step. This step transforms the filtered projection data to
the image domain. Using the notation of \eqref{eq:IP}, the FDK algorithm is
given by
\begin{align}
  \text{FDK}(\mathbf{y}, \mathbf{h}) = W^T(\mathbf{h} * r\brac{\mathbf{y}})_{\text{1D}},\label{eq:FDK2}
\end{align}
with $W^T$ the transpose of $W$. The operator $W^T$ is also known as the \emph{backprojection operator}.

In \cite{pelt2013fast, pelt2014improving, lagerwerf2020automated} exponential
binning is used to approximate filters, leading to $N_e\approx\log{N}$ coefficients to describe a filter. This approximation can be
seen as a matrix $E\in \R^{2N\times N_e}$ applied to a coefficient vector
$\h_e\in\R^{N_e}$:
\begin{align}
  \h \approx E \h_e.
\end{align}
The implementation details of this filter approximation can be found in
\cite{lagerwerf2020automated}.

\subsubsection{Perceptron}
In a similar manner as in \cite{bishop2006pattern} we define a \emph{perceptron}
or \emph{node} $\mathsf{P}:\R^l\rightarrow \R$ as a non-linear activation
function $\sigma:\R\rightarrow\R$ applied to a weighted sum of the input $\eta
\in \R^l$ with the weights $\xi \in \R^l$ and a bias $b\in\R$:
\begin{align}
  \mathsf{P}_{\xi, b}(\eta)=\sigma(\eta \cdot \xi - b)\label{eq:perceptron}
\end{align}
In this paper we will only consider the sigmoid function as activation function,
\ie $\sigma(t)=1/(1+e^{-t})$.

A multilayer perceptron is a network structure containing two types of layers
with perceptrons, where each perceptron operates on the outputs of the previous
layer. These layers are, in order, any number of
\emph{hidden layers}, and the \emph{output layer}. Note that the number of
hidden layers and number of hidden nodes $N_h$ in these layers can be chosen
freely.

\subsection{Reconstruction algorithm \& Network design}\label{sec:deriv}
We formulate the NN-FDK reconstruction algorithm in a similar fashion as the
NN-FBP method in
\cite{pelt2013fast}. 
The NN-FDK reconstruction algorithm consists of $N_h$ individual FDK algorithms
executed on the input data $y$, each using its own (exponentially binned) filter
$\h_e^k \in \R^{N_e}$. It combines these $N_h$ volumes into a single
reconstruction, using point-wise application of the activation function $\sigma$
and an output perceptron with parameters $b_o, b_k \in \R$, and $\xi \in
\R^{N_h}$.


We use $\theta = ( \xi, b_o, \h_e^k, b_k )$ as short-hand for the full set of
parameters of the NN-FDK reconstruction algorithm. The full algorithm is then
given by the following equation.

\begin{align}
\text{NN-FDK}_\theta(\y)= \sigma\Big( \sum^{N_h}_{k=1}\xi_k\sigma\brac{\FDK(\y, E\h_e^k)-b_k} - b_o\Big)\label{eq:NNFDK}
\end{align}
The FDK algorithm is a bilinear map in the input projection data and the used
filter. Therefore, for fixed input projection data $\y$ and an expanded
exponentially binned filter $E \h_e$, the FDK algorithm can be written as a
linear map $\Fy$ applied to $E \h_e$. The product $\Fy E$ can be considered as a
matrix of size $N^3 \times N_{e}$, and the $v$-th voxel of the output of the FDK
algorithm is given by the inner product of $\h_e$ with $(\Fy E)_{v:}$, the
$v$-th row of the matrix $\Fy E$. This leads to the following:
\begin{align}
  (\text{NN-FDK}_\theta(\y))_v&= \sigma\Big( \sum^{N_h}_{k=1}\xi_k\sigma\brac{(\Fy E\h_e^k)_v-b_k} - b_o\Big),\\
                               &=  \sigma\Big( \sum^{N_h}_{k=1}\xi_k\sigma\brac{(\Fy E)_{v:}\h_e^k-b_k} - b_o\Big),\\
  & = \mathsf{P}_{\xi, b_o}\left( \left[ \mathsf{P}_{\h_e^k, b_k}((\Fy E)_{v:} ) \right]_k \right).
\end{align}
Therefore, we define the two-layer perceptron network $\mathsf{N}_\theta : \R^{N_e} \to
\R$:
\begin{align}
  \mathsf{N}_\theta(\q) = \mathsf{P}_{\xi, b_0} \left( \left[ \mathsf{P}_{\h_e^k, b_k} (\q) \right]_k \right).
\end{align}
This is our NN-FDK network, and as we derived above, it has the following
relationship with the NN-FDK reconstruction algorithm:
\begin{align}
\mathsf{N}_\theta((\Fy E)_{v:}) = (\text{NN-FDK}_\theta(\y))_v.\label{eq:net_rel}
\end{align}
This relationship shows that we can \emph{evaluate} the NN-FDK reconstruction
algorithm efficiently on full input projection data at once, but also
\emph{train} the NN-FDK network efficiently with each \emph{individual} voxel
$(\mathbf{x}_{\text{HQ}})_v$ in a high quality reconstruction yielding a
training pair with input $(\Fy E)_{v:}$ and target $(\mathbf{x}_{\text{HQ}})_v$.
A schematic representation of the network is given in \figref{fig:scheme}.

Note that we arrive at the same network structure as found in
\cite{pelt2013fast} for FBP, using only the properties that the FDK algorithm is
a bilinear map in the data and the filter, and that all operations can be
applied point-wise. Using this reasoning we can derive a similar network
structure for any FBP-type method satisfying these conditions.


Even though we use the same network structure as \cite{pelt2013fast}, the way we compute inputs to the network is different. In \cite{pelt2013fast}, the input to the NN-FBP network is explicitly calculated by shifting and adding projection data for each reconstruction pixel. The FDK algorithm has additional weighting factors and lacks the shift-invariance property, which makes the approach presented in \cite{pelt2013fast} not directly applicable. In the next section, we detail
an alternative method to compute the input. The same approach could be applied to the NN-FBP method, similarly simplifying the network input computations.  

\begin{algorithm}[!t]
  \caption{Neural Network FDK reconstruction algorithm}
  \begin{algorithmic}[1]\label{alg:NNFDK}
    \STATE{Given a set of parameters, ${\theta}:=\brac{\xi, b_o, \mathbf{h}^k_e,
        {b}_k}$.} \STATE{Compute $H_k$ for all nodes $k$ of
      the hidden layer:}\\
    \FOR{$k = \{1,2,..,N_h\}$}
    \STATE{$\text{H}_k(y)=\sigma\brac{\FDK(\mathbf{y}, E\mathbf{h}_e^k)-{b}_k}$}
    \ENDFOR
    \STATE{Compute the output of the output layer:\\
      $\text{NN-FDK}_{\theta}(\mathbf{y})=\sigma\brac{\sum_{k=1}^{N_h}\mathbf{\xi}_kH_k(\mathbf{y})-{b}_o}$}
  \end{algorithmic}
\end{algorithm}

\tikzstyle{object} = [rectangle, draw, line width=1pt, fill=orange!25, text
width=1.75em, node distance=2.4cm, drop shadow]

\tikzstyle{compute} = [rectangle,draw, fill=orange!65, line width=1pt, text
width=6em, text centered, rounded corners, minimum height=3em, drop shadow, node
distance=1.2cm]

\tikzstyle{line} = [draw, -latex', line width=1.25pt, ]

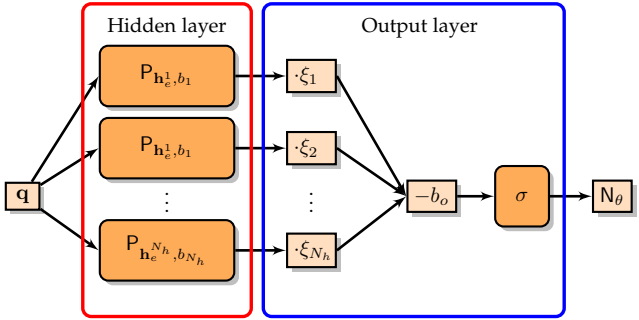
\begin{figure}[t]
  \centering \scalebox{.8}{\parbox{1.2\linewidth}{%
  \begin{tikzpicture}[node distance = 2cm, auto]
  \node  (dots) {$\vdots$}; 
  \node [object, text width=1em, text centered, left of=dots] (input) {$\mathbf{q}$};
  \node [compute, above of=dots, node distance=.8cm] (FDK2) {$\mathsf{P}_{\mathbf{h}^1_e, b_1}$};
  \node [compute, above of=FDK2] (FDK1) {$\mathsf{P}_{\mathbf{h}^1_e, b_1}$};
  \node [compute, below of=dots, node distance=.9cm] (FDK3) {$\mathsf{P}_{\mathbf{h}^{N_h}_e, b_{N_h}}$};
  \node [object, right of=FDK1] (xi1) {$\cdot\xi_1$};
  \node [right of=dots, node distance=2.4cm] (dots2) {$\vdots$}; 
  \node [object, right of=FDK2] (xi2) {$\cdot\xi_2$};
  \node [object, right of=FDK3] (xi3) {$\cdot\xi_{N_h}$};
  \node [object, right of=dots2, node distance=2cm] (bias) {$-b_o$};
  \node [compute, right of=bias, text width=2em, node distance=1.5cm] (sigma) {$\sigma$};
  \node [object, right of=sigma, node distance=1.5cm, text width=1.25em] (Network) {$\mathsf{N}_\theta$};
  \path [line] (input) -\ (FDK1.west); \path [line] (input) -\ (FDK2.west);
  \path [line] (input) -/ (FDK3.west); \path [line] (FDK1.east) -- (xi1.west);
  \path [line] (FDK2.east) -- (xi2.west); \path [line] (FDK3.east) --
  (xi3.west); \path [line] (xi1.east) -- (bias.west); \path [line] (xi2.east) --
  (bias.west); \path [line] (xi3.east) -- (bias.west); \path [line] (bias.east)
  -- (sigma.west); \path [line] (sigma.east) -- (Network.west);
  \draw[red,ultra thick, rounded corners] (-1.4, 3.2) rectangle (1.4, -2);
  \draw[blue,ultra thick, rounded corners](1.6, 3.2) rectangle (6.6, -2);
  \node at (0, 2.8) {Hidden layer};
  \node at (4.2, 2.8) {Output layer};
\end{tikzpicture}
\caption{\large{Schematic representation of the NN-FDK network,
    $\mathsf{N}_\theta:\R^{N_e}\rightarrow\R$, with $N_h$ hidden nodes. Note
    that if we take $q=(\Fy E)_{v:}$ we get $q\cdot \h_e^k = (\FDK(\y,
    E\h_e^k))_v$ in the perceptrons of the hidden layer and the output of the
    network is equal to the $v$-th voxel of the NN-FDK reconstruction algorithm.}}
\label{fig:scheme}}}
\end{figure}

\subsection{Training process}\label{sec:training}
\subsubsection{Training and validation data}
We will train our network using supervised learning, where we assume that we
have $N_\text{TD}$ and $N_\text{VD}$ datasets available for training and
validation, respectively. These datasets consist of low quality tomographic
input data and a high quality reconstruction from which we randomly draw a total
of $N_\text{T}$ training pairs and $N_\text{V}$ validation pairs. Note that we ensure that
every drawn pair is unique and that an equal number of pairs is taken from each
dataset. Moreover, to avoid selecting too many training pairs from the
background we only take training pairs from a region of interest (ROI) around
the scanned object. This ROI is defined from the high quality reconstruction as
the voxels in the reconstructed object plus a buffer of roughly $0.2N$ voxels
around it.



Recall from the previous section that given low quality tomographic data $\y$
and a high quality reconstruction $\mathbf{x}_\text{HQ}$ the matrix $\Fy E$ contains each 
input vector $Z=\brac{\Fy E}_{v:}\in \R^{N_e}$ corresponding to the target voxel $O=(\mathbf{x}_\text{HQ})_v$. However, due to memory constraints $\Fy E$ cannot be computed directly as a matrix product. Therefore, we observe that each column of $\Fy E$ is an FDK reconstruction with a specific filter:
\begin{align}
  (\Fy E)_{:j} = \Fy E \mathbf{e}_j = \text{FDK}(\mathbf{y}, E\mathbf{e}_j), 
\end{align}
with $\mathbf{e}_j\in \R^{N_e}$ the unit vector with all entries equal to zero
except for the $j$-th element.

\subsubsection{Learning problem}\label{sec:LP}
The parameters of the NN-FDK network are learned by finding the set of
parameters $\theta^\star$ that minimize the \emph{loss function} $\mathcal{L}$
on the {training set}. We minimize the $\ell^2$-distance between the network
output and the target voxel for all training pairs in $T$:
\begin{align}
  \theta^\star&=\argmin_{\theta}\mathcal{L}(\mathbf{\theta},T) =\argmin_{\theta}\tfrac{1}{2}\sum^{N_{\text{T}}}_{j=1}\brac{O_j-\mathsf{N}_{\theta}(Z_j)}^2.\label{eq:MLP_MP}            
\end{align}
To minimize the loss function we use a quasi-Newton optimization scheme, the
\emph{Levenberg-Marquardt algorithm} (LMA) as proposed in
\cite{levenberg1944method, marquardt1963algorithm}. This is a combination of
gradient descent and the Gauss-Newton algorithm, improving the stability of
Gauss-Newton while retaining its fast convergence and it is specifically
designed to minimize a non-linear least squares problem such as
\eqref{eq:MLP_MP}. Note that the small number of parameters of the proposed
network allows us to use such a method. Lastly, to avoid overfitting we check
whether every update of the parameters also reduces the loss function on the validation
set. We discuss the specifics of this algorithm in \appref{sec:LMA}.

\subsection{Method characteristics \& comparison}\label{sec:theo_comp}
To conclude the method section we compare the characteristics of the NN-FDK algorithm
to those of several other methods. These methods are two 2D
post-processing DNNs (U-net \cite{ronneberger2015u} and MSD-net
\cite{pelt2018mixed}) applied in a slice-by-slice fashion, the SIRT$^{+}$
algorithm \cite{van1990sirt} and the FDK algorithm. We focus our discussion on
the goals formulated in \secref{sec:intro} and show a summary of this comparison
in \tabref{tab:comp}. The reconstruction accuracy will be discussed in
\secref{sec:results}.
\begin{table}[!tp]
  \centering
  \begin{tabular}{|l|c|c|c|c|}
\multicolumn{5}{c}{\normalsize{Method comparison: Goals}}\\
\specialrule{.2em}{.05em}{.05em} 
     \multicolumn{1}{c}{} & \multicolumn{2}{c}{Reconstruction} & \multicolumn{2}{c}{Training} \\
    \hline
    Method & Time & Accuracy & Data & Time \\
    \hline
    \hline
    NN-FDK & ++  & ?    &  ++ & +++ \\
    DNN & $\pm$   & +++  & $\pm$   & - - - \\
    \hline
    FDK    & +++ & - -   &   &  \\
    SIRT$^{+}$ & - - & +   &  & \\
    \hline   
  \end{tabular}
  \caption{Comparison of reconstruction methods with respect to the goals
    formulated in \secref{sec:intro}. We consider a DNN to be 2D deep
    convolutional neural network (U-net \& MSD-net) applied in slice-by-slice
    fashion to a standard FDK reconstruction. Reconstruction accuracy is defined
    as the accuracy of a method when reconstructing low quality data, \eg data
    with high noise or a low number of projection angles. }
  \label{tab:comp}
\end{table}
\subsubsection{Computational efficiency}
We approximate the reconstruction time by counting how many
times it has to evaluate its most expensive computations. For simplicity we
assume that a backprojection takes approximately the same time as a forward
projection, $T_{\text{BP}}$.
\begin{itemize}
  \item \textbf{FDK:} The FDK algorithm consist of one reweighting, filtering and
    backprojection step, \ie:
    \begin{align}
      T_{\text{FDK}} \approx T_{\text{BP}}.
    \end{align}
  \item \textbf{NN-FDK:} The NN-FDK algorithm performs one FDK
    reconstruction per hidden node $N_h$. Therefore the reconstruction time
    becomes:
    \begin{align}
      T_{\text{NN-FDK}} \approx N_h T_{\text{BP}}.
    \end{align}
  \item \textbf{SIRT$^+$:} The SIRT$^{+}$ method evaluates a forward and
    backprojection for each iteration. For $N_{\text{iter}}$ iterations, the
    reconstruction time becomes:
    \begin{align}
      T_{\text{SIRT}^+}\approx 2N_{\text{iter}}T_{\text{BP}}.
    \end{align}
  \item \textbf{DNN:} To evaluate a DNN an FDK reconstruction is performed and a
    2D network is applied per slice of the FDK reconstruction.
    \begin{align}
  T_{\text{DNN}}\approx T_{\text{BP}} + N T_{\text{DNN}},
    \end{align}
    with $T_{\text{DNN}}$ the time it takes to apply a 2D DNN.
\end{itemize}
On a modern GPU and with $N=1024$ and $N_a=360$, we found in our experiments that
$T_\text{BP}\approx 10$ s and $T_\text{DNN}\approx 0.5$ s.

Comparing the reconstruction times, we see that NN-FDK is similar to FDK when
the number of nodes $N_h$ is small, which is the case since we will take
$N_h{=}4$ (see \secref{sec:params}). For DNNs the computational load of applying
a 2D network leads to relatively high reconstruction times compared to the FDK
algorithm. Lastly, we note that the number of iterations $N_{\text{iter}}$ often
lies between the 20 and 200, making SIRT$^+$ several times slower than the
(NN-)FDK algorithm.

\subsubsection{Number of trainable parameters}
The number of trainable parameters is closely related to the amount of training
data required to train a network \cite{pelt2018mixed}. From the definition of
the NN-FDK network \eqref{eq:NNFDK} we can compute the number of trainable
parameters $\abs{\theta}$:
\begin{align}
  \abs{\theta}= (N_e + 2)N_h + 1,
\end{align}
with $N\gg N_h,N_e>0$. Taking $N_h=4$ and $N=1024$ gives
$\abs{\theta}=61$, which is several orders of magnitude lower than the typical
numbers of parameters in a DNN (several tens of thousands to millions). 

\subsubsection{Training time}\label{sec:theo_train_time}
In the training step a solution to the minimization problem \eqref{eq:MLP_MP} is
computed. For the NN-FDK algorithm this problem has $N_{\text{T}}$ samples and
$\abs{\theta}$ unknowns. In a similar fashion we can formulate a least squares
problem for training a DNN.
Even assuming that we only take the same number of training
samples to train the DNNs, this least squares problem is already orders of magnitude larger than that for NN-FDK
due to the difference in the number of trainable parameters.
Moreover, the LMA
(the algorithm used to train NN-FDK) approaches quadratic convergence, which
means it will need fewer iterations to converge than a first order scheme such
as ADAM \cite{kingma2014adam}, which is often used for training DNNs.
Considering these two observations we expect the training time of the NN-FDK
algorithm to be lower than the training time of the DNNs.


\section{Experimental setup}\label{sec:experiments}
We carried out a range of experiments to assess the performance of the NN-FDK
algorithm with respect to the goals formulated in \secref{sec:intro} compared to
several alternative methods. In this section we introduce the setup of these
experiments. We describe the simulated data in \secref{sec:sim_data} and the
experimental data in \secref{sec:exp_data}. In \secref{sec:params} we discuss
the specific network structure for the NN-FDK algorithm and the training
parameters used. Finally, we give the quantitative measures we use to compare
the reconstruction in \secref{sec:QM}.

\subsection{Simulated data}\label{sec:sim_data}
\begin{figure}[!h]
  \includegraphics[width=0.48\textwidth]{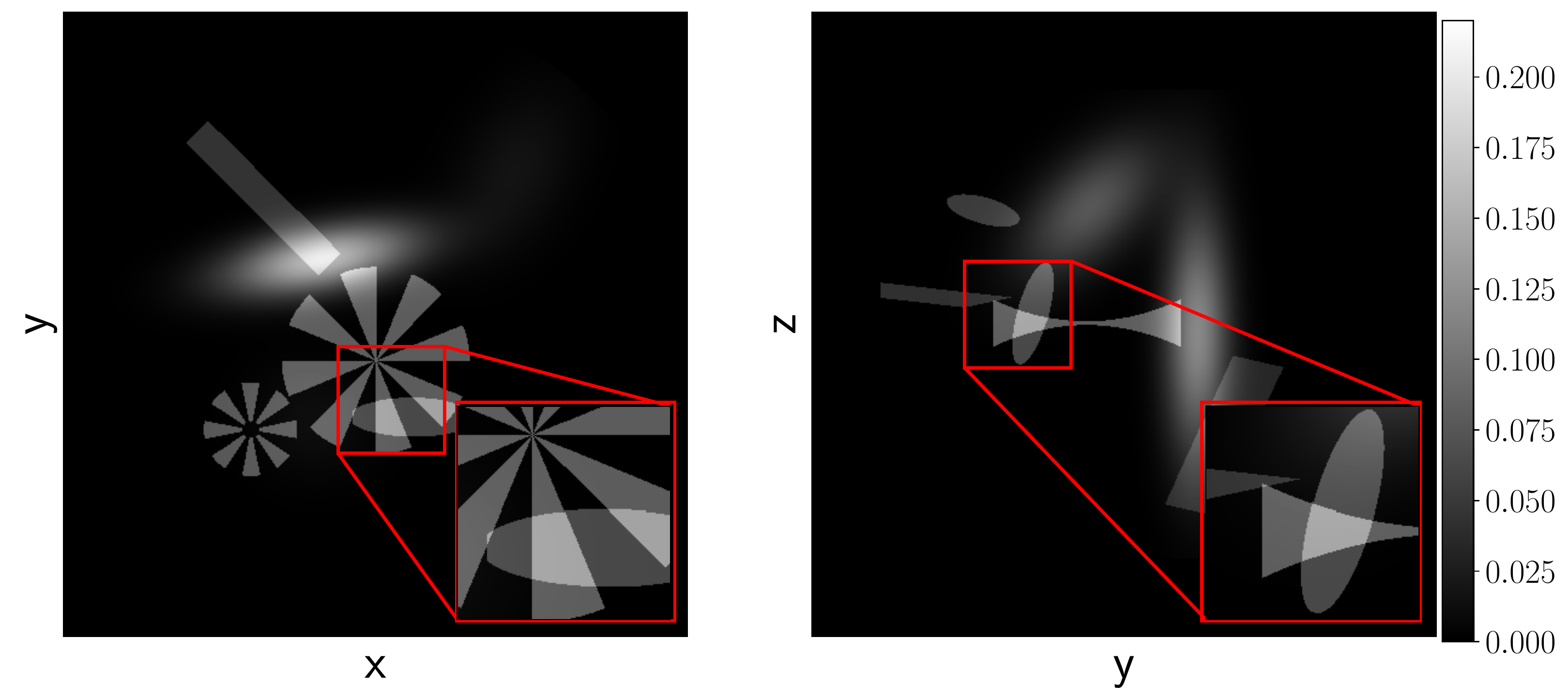}
  \caption{Slices, (Left) $z=0$, (Right) $x=0$, of the Fourshape test phantom.
    This phantom is designed such that at least one of all objects can clearly
    be observed in the slices.}
\label{fig:4S}
\includegraphics[width=0.48\textwidth]{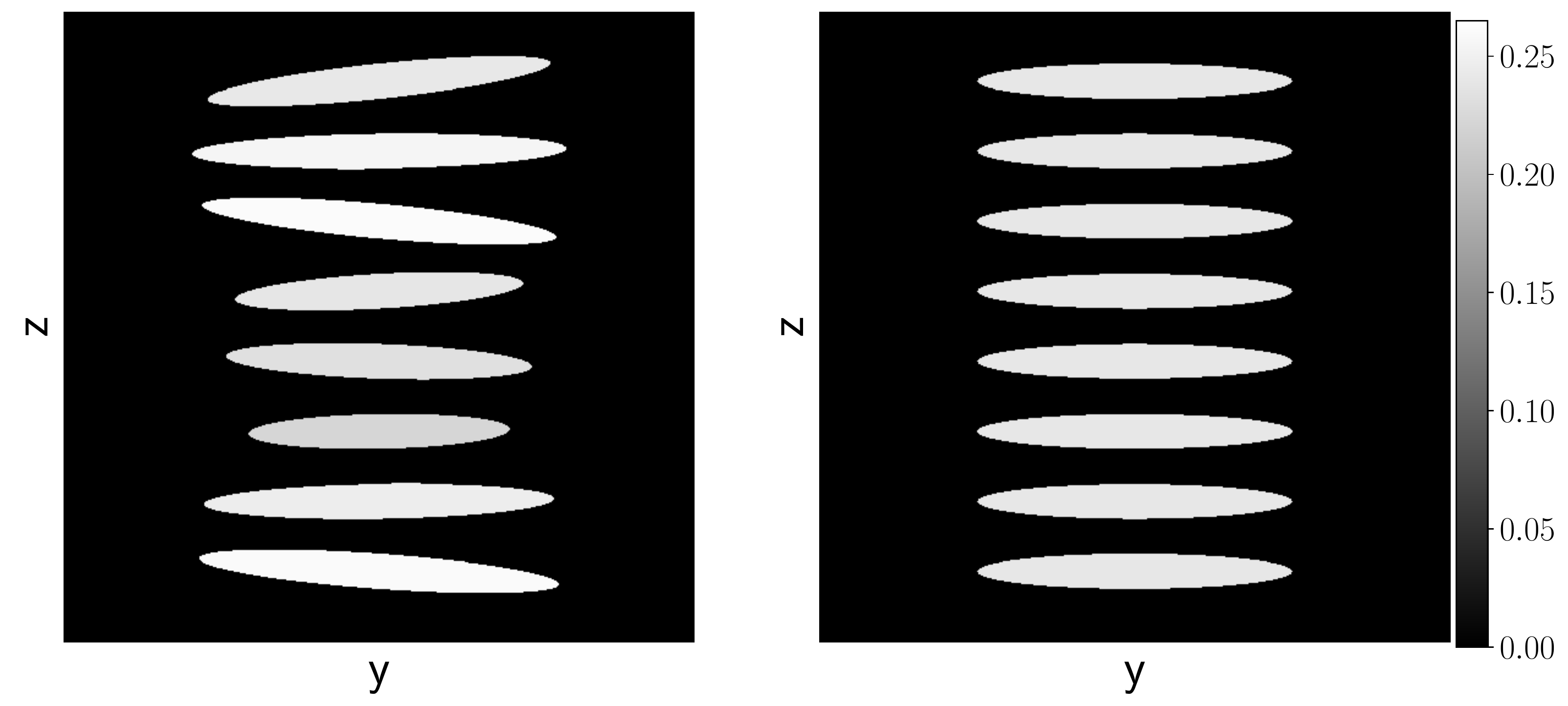}
\caption{The $x=0$ slice for a Random Defrise phantom (Left) and the standard
  Defrise phantom without alternating intensities from \cite{kudo1998cone}
  (Right).}\label{fig:DF}
\end{figure}

We consider two types of phantom families for the simulated data experiments:
the \emph{Fourshape phantom family} and the \emph{Random Defrise phantom
  family}. Examples are shown in \figref{fig:4S} and \figref{fig:DF},
respectively. The Fourshape phantom family contains three random occurrences of
each of four types of objects: an ellipse, a rectangle, a Gaussian blob and a Siemens
star. For evaluation and visualization of the reconstructions we fixed one realization that
clearly shows at least one of all the four objects and we will refer to this
phantom as the \emph{Fourshape test phantom}. The Random Defrise phantom family
is a slight adaptation of the phantom introduced in \cite{kudo1998cone}, which
is a common phantom for assessing the influence of imaging artifacts due to the
cone angle. Here we vary the intensities, orientations and sizes of the disks
making sure they do not overlap. Again, we define a test phantom for evaluation and 
visualization, which is in this case the standard \emph{Defrise phantom} without
alternating intensities (right in \figref{fig:DF}). To simulate realistic
settings, we scale the phantoms to fit inside a 10 cm cube, and use an
attenuation coefficient of $\mu=0.22$ cm$^{-1}$, approximating that of various common
plastics at 40 keV \cite{hubbell1995tables}. These phantoms are defined through
geometric parameters, and can therefore be generated for any desired $N$. For
our experiments we will take $N=1024$. Details about how we generate the data
are given in \appref{sec:data_gen}.

To compute a high quality reconstruction $\mathbf{x}_{\text{HQ}}$ that can be
used as target for training (recall \secref{sec:training}) we consider a
simulated dataset with $N_a=1500$ projection angles, low noise ($I_0=2^{20}$
emitted photon count) and cone angle of $0.6$ degrees and reconstruct this
problem with the standard FDK algorithm using a Hann filter
\cite{natterer2001mathematics}.

\subsection{Experimental data}\label{sec:exp_data}
For experimental data we consider a set of CT scans that were recorded using
the custom-built and highly flexible FleX-ray CT scanner, developed by XRE NV
and located at CWI \cite{coban2020explorative}. This scanner has a flat panel
detector with $972 \times 768$ pixels and a physical size of $145.34 \times
114.82$ mm. This set of 42 scans was set up to create high noise
reconstruction problems and low noise reconstruction problems with a low number
of projection angles. 

\begin{figure}[!h]
  \centering \includegraphics[width=.48\textwidth]{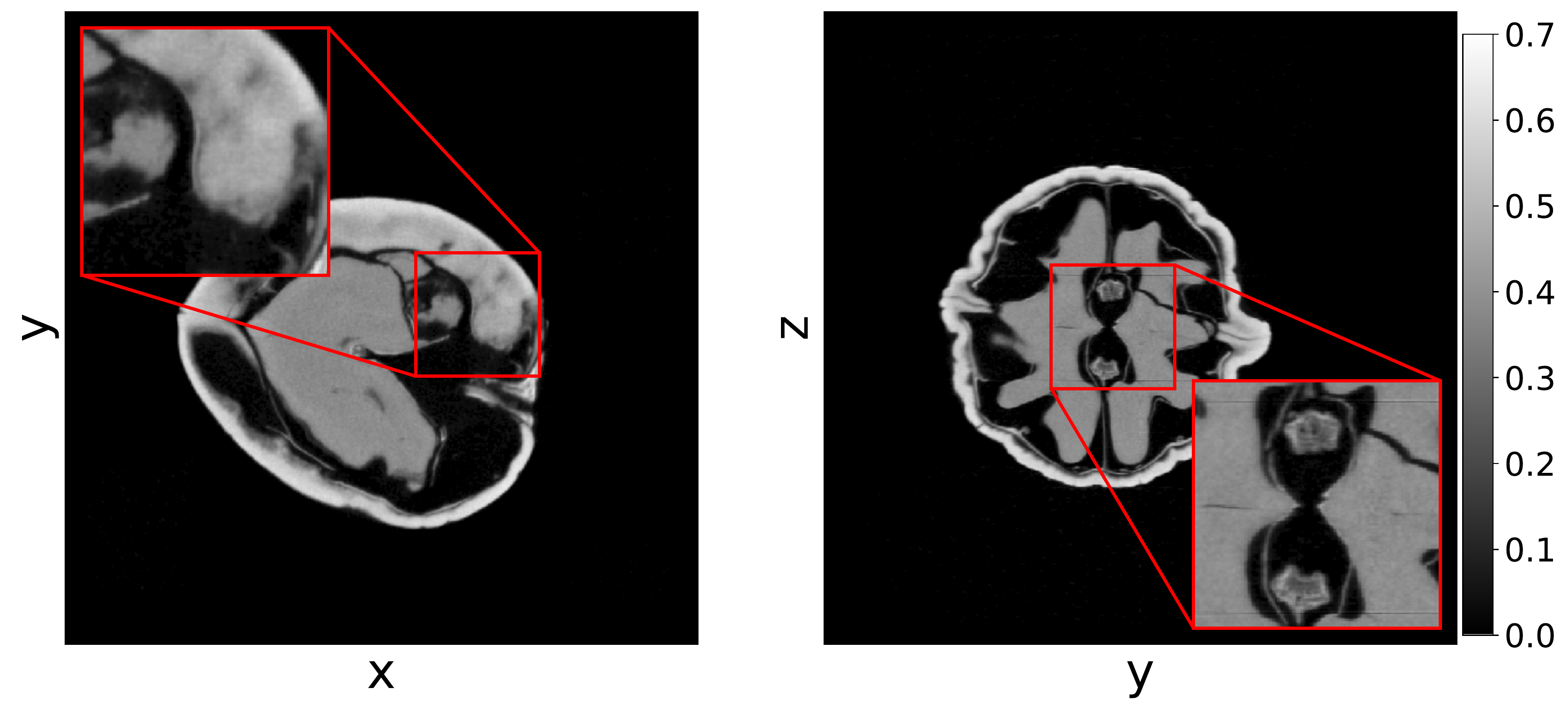}
  \caption{The $z=0$ (Left) and $y=0$ (Right) slice of the gold standard
    reconstruction of the high-dose dataset of the $21^{st}$ walnut with full
    number of projection angles. The projection data is acquired using the
    FleX-ray scanner located at the CWI \cite{walnuts}.}
  \label{fig:real_data}
\end{figure}

We acquired high-dose (low noise) and low-dose (high noise) scans of 21 walnuts. The
datasets contain 500 equidistantly spaced projections over a full circle. The
distance from the center of rotation to the detector was set to 376 mm and the
distance from the source to the center of rotation was set to 463 mm. The scans
were performed with a tube voltage of 70 kV. The high-dose scan was collected
with a tube power of 45 W and an exposure time of 500 ms per projection. The
low-dose scan was collected with a tube power of 20 W and an exposure time of
100 ms per projection. To create a low noise reconstruction problem with a low
number of projection angles we considered the high-dose scan but only took every
16-th projection angle. As high quality reference reconstructions we used
SIRT$^{+}$ reconstructions with 300 iterations (SIRT$^+_{300}$) of the high-dose
scans with all available projection angles ($N_a=500$). We will refer to these
reconstructions as the \emph{gold standard} reconstruction and we show such a
reconstruction in \figref{fig:real_data}. These datasets are available at Zenodo
\cite{walnuts}.

\subsection{Parameter settings NN-FDK}\label{sec:params}
\subsubsection{Network structure}\label{sec:network_struc}
In our initial experiments we found that taking more FDK-perceptrons improved
the accuracy of the networks, at the cost of increasing the training and
reconstruction time. We found that $N_h{=}\ 4$ FDK-perceptrons led to a good
balance between accuracy and reconstruction time, which is similar to the
findings in \cite{pelt2013fast}.

\subsubsection{Training data}\label{sec:train_data}
We found that, similar to the findings in \cite{pelt2013fast}, taking
$N_{\text{T}}=10^6$ voxels for training and $N_{\text{V}}=10^6$ for validation is sufficient
for training an NN-FDK network. 

The network structures and training procedure used for the U-nets and MSD networks are discussed in \appref{sec:DNNs}.

\subsection{Quantitative measures}\label{sec:QM}
To quantify the accuracy of the reconstructions we consider two measures, the
test set error (TSE) and the structural similarity index (SSIM). These measures
compare the reconstructed image $\mathbf{x}_{r}$ to a high quality
reconstruction $\mathbf{x}_{\text{HQ}}$ on the ROI (as
discussed in \secref{sec:training}).

The TSE is the average loss\footnote{Recall \eqref{eq:MLP_MP} in
  \secref{sec:training}} of the test set, where the test set is all the voxels
defined in the ROI of $\mathbf{x}_{\text{HQ}}$:
\begin{align}
  \text{TSE}(\mathbf{x}_r, \mathbf{x}_{\text{HQ}})&=\tfrac{1}{N_{\text{ROI}}}\mathcal{L}(\mathcal{I}_{\text{ROI}}(\mathbf{x}_{\text{HQ}}), \theta),\\
                                                  &= \tfrac{1}{2N_{\text{ROI}}}\norm{\mathcal{I}_{\text{ROI}}(\mathbf{x}_{\text{HQ}}-\mathbf{x}_r)}^2_2.
\end{align}
with $\mathcal{I}_\text{ROI}:\R^{N^3}\rightarrow \R^{N^3}$ the masking function for the
ROI and $N_{\text{ROI}}$ the number of voxels in the ROI.

The SSIM \cite{wang2004image} is implemented based on the scikit-image 0.13.1
\cite{scikit-image} package, where all the constants are set to default and the
filter is uniform with a width of 19 pixels. 

\section{Results and discussion}\label{sec:results}
\subsection{Scalability}
\subsubsection{Memory scaling}\label{sec:mem_con}
\begin{figure}
    \centering
    \includegraphics[width=0.5\textwidth]{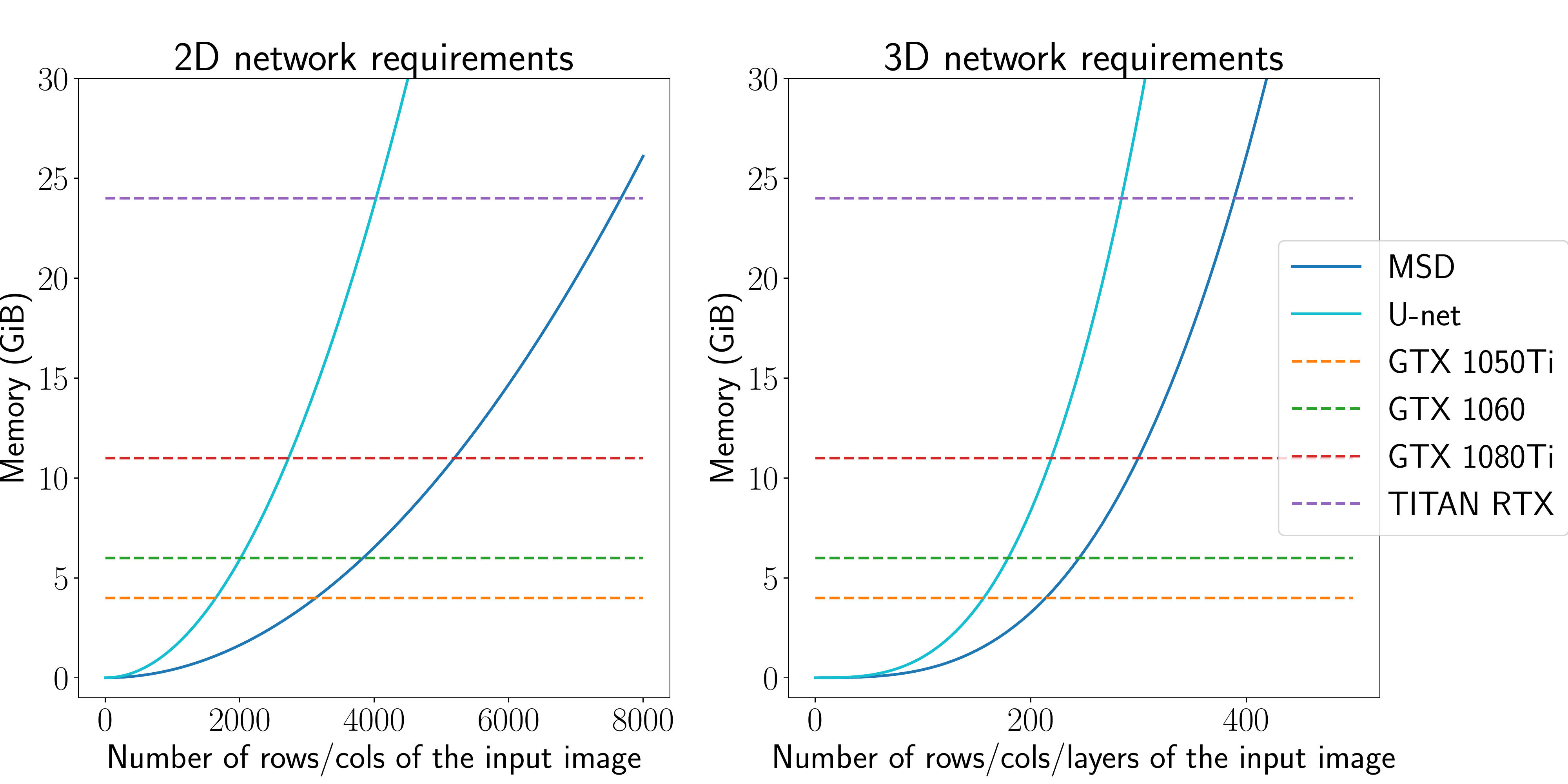}
    \caption{The required memory to store all intermediate images for applying a 2D and 3D U-net and MSD network as a function of the input image size.}
    \label{fig:mem_con}
\end{figure}

The required memory to store all intermediate images for a forward pass of a 2D or a 3D U-net and MSD network as a function of the input image size is shown in \figref{fig:mem_con}. Considering that CT imaging problems typically range from $256\times 256\times 256$ up to $4096\times 4096 \times 4096$ we conclude from these figures that full 3D networks do not fit into GPU memory for higher resolutions and that even for 2D U-nets not all resolutions fit on the GPU. As a forward pass of the NN-FDK algorithm requires only one additional reconstruction volume\footnote{Technically a forward pass of the NN-FDK algorithm can be done for every voxel separately, however, for the sake of comparison we assume a forward pass is for a full reconstruction volume.} compared to the FDK algorithm,  the memory requirements of the NN-FDK algorithm are roughly 2 times the memory required by the FDK algorithm.
\subsubsection{Training time}\label{sec:train_time}
\begin{figure}
    \centering
    \includegraphics[width=0.5\textwidth]{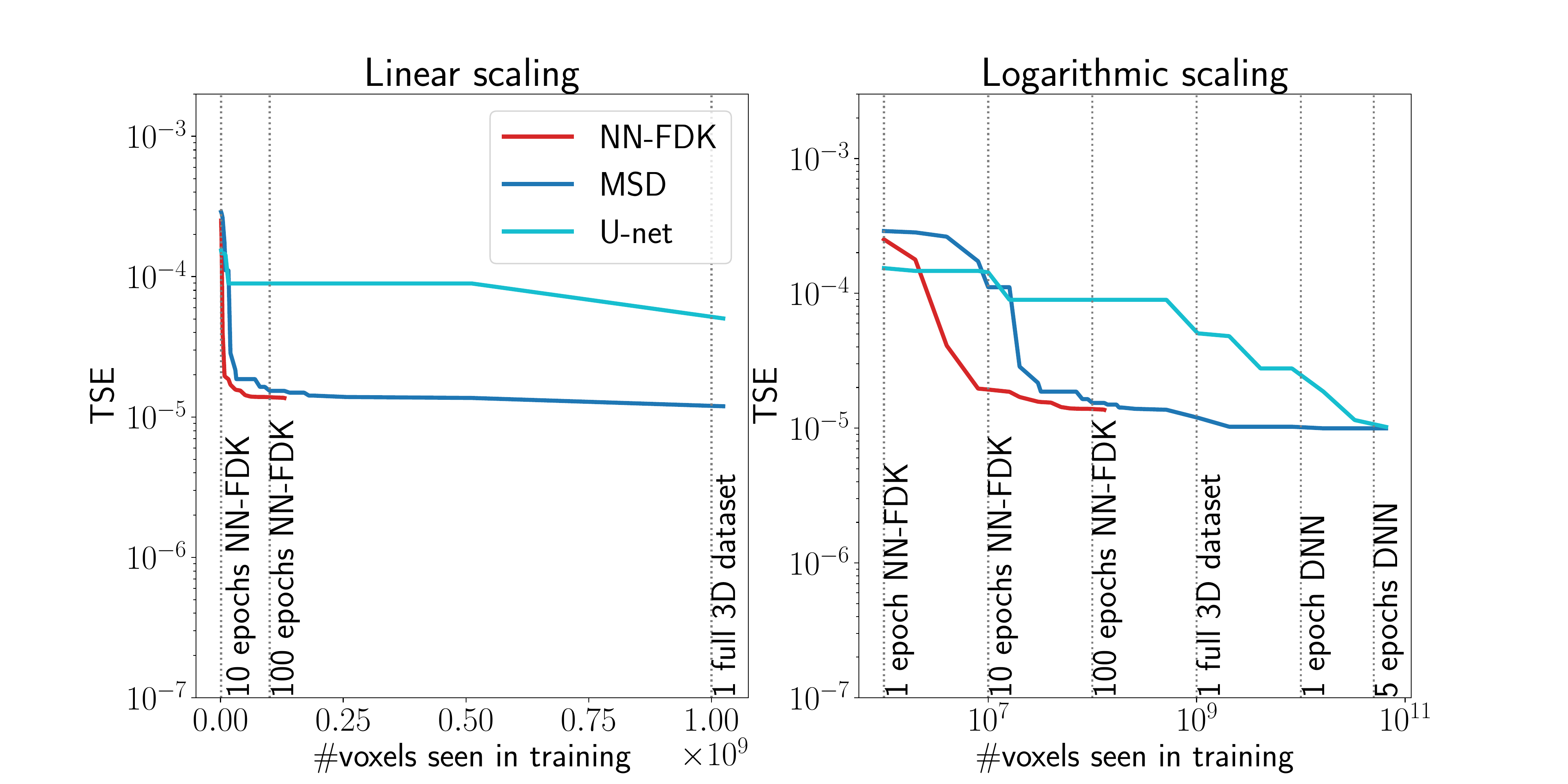}
    \caption{The TSE as a function of the number of voxels the training process has seen. We report the lowest TSE up till that point. The networks are trained on randomly generated Fourshape phantoms with size $N=1024$, $N_a=32$ projection angles and no noise.  (Left) Linear scaling in the number of voxels ranging from 1 epoch for the NN-FDK ($10^6$ voxels), to 1 full 3D dataset ($10^9$ voxels). (Right) Logarithmic scaling in the number of voxels. Ranging from 1 epoch for the NN-FDK network ($10^6$ voxels) to 5 epochs for a DNN ($5\cdot 10^{10}$ voxels). }
    \label{fig:epochs}
\end{figure}

In \figref{fig:epochs} we compare the training processes by plotting the progress of the network training (measured by the TSE) as a function of the number of voxels that the network has seen during training. 
We see that the NN-FDK has seen $1.1\cdot 10^8$ voxels when it converges to TSE$=1.4\cdot10^{-5}$, whereas, MSD and U-net have seen $5.1\cdot 10^8$ voxels and $3.2\cdot 10^9$ voxels, respectively, at the point they first achieve a similar TSE. Important to note is that both U-net and MSD are not yet converged when they match the TSE of NN-FDK, and in general the DNNs achieve lower TSEs than NN-FDK. 

In \tabref{tab:train_time} we show various timings and properties with respect to the training process. These timings are recorded using one Nvidia GeForce GTX 1080Ti with 11GiB memory. We define a converged training process as 100 epochs without improvement on the validation set error and the number of epochs to converge as the epoch with the lowest validation set error during a converged training process. From these results we see that the size of the training problem influences the time per epoch as an NN-FDK epoch is sub-second and the time per epoch for DNNs is in the range of hours. 

In practice, we observed that after 2 days of training for the DNNs, any additional training only achieved marginal improvements. Therefore, in the following experiments we train all DNNs for 2 days with one Nvidia GeForce GTX 1080Ti GPU, unless mentioned otherwise. 

\begin{table}[!h]
  \centering 
  \begin{tabular}{|l|r|r|r|}
  \multicolumn{4}{c}{\normalsize{Training process}}\\
  \specialrule{.2em}{.05em}{.05em}
  & NN-FDK$_4$ & MSD & U-net \\
    \hline
Voxels seen in one epoch & $1 \cdot 10^6$ & $1.1 \cdot 10^{10}$ & $1.1 \cdot 10^{10}$\\ 
  Time per epoch    & 0.1336 (s) & 0.95 (h) & 2.36 (h) \\
  Time to converge  & 28 (s) & $\pm$ 10 (d) & $\pm$ 14 (d) \\
  Epochs to converge& 110 & 128 & 42  \\
  Epochs in 2 days  & - & 45 & 18 \\
  \hline
  \end{tabular}
  \caption{Timings and properties of the considered training processes. We define a converged training process as 100 epochs without improvement on the validation set error. The epochs to converge is therefore the epochs computed of such a process minus 100. The training was performed using one Nvidia GeForce GTX 1080Ti GPU (11 GiB).}
  \label{tab:train_time}
\end{table}

\subsubsection{Reconstruction time}
We measured the average reconstruction times and corresponding standard deviation over 120 reconstructions with resolution $N^3=1024^3$ and $N_a=360$ projection angles. These reconstructions are computed using one Nvidia GeForce GTX 1080Ti with 11 GiB memory. The results are shown in \tabref{tab:rec_time}. We define the reconstruction time as the time it takes to compute the full 3D volume. This means for U-net and MSD, an FDK reconstruction needs to be computed and the network needs to be applied $N=1024$ times to a 2D slice. Although every application can be done within a second (U-net $\approx0.3 s$, MSD $\approx0.7 s$) this leads to long reconstruction times. 

\begin{table}[!htp]
  \centering
  \begin{tabular}{|r|r|r|r|r|}
    \multicolumn{5}{c}{\normalsize{Reconstruction times}}\\
    \specialrule{.2em}{.05em}{.05em}
     FDK & SIRT$^+_{200}$ & NN-FDK$_{4}$ & U-net & MSD \\
    \hline
 \textbf{28 $\pm$ 8} & 3225 $\pm$ 916 & 76 $\pm$ 8 &382 $\pm$ 69 & 809 $\pm$ 86\\
    \hline
\end{tabular}
\caption{Average and standard deviation of the reconstruction times (in seconds) computed
   over 120 reconstruction problems with $N=1024$ and $N_a=360$ projection angles. These reconstructions are computed using one Nvidia GeForce GTX 1080Ti GPU (11 GiB).}
\label{tab:rec_time}
\end{table}



\subsection{Reconstruction accuracy for simulated data}\label{sec:rec_acc_sim}
For evaluating the reconstruction accuracy using simulated data, we consider 16 cases: 6 different noise levels, 5 different numbers of projection angles and 5 different cone angles. For each case an NN-FDK, MSD and U-net network was trained. For the training process of NN-FDK we used $N_\text{T}=10^6$ training voxels and $N_\text{V}=10^6$ validation voxels from $N_\text{TD}=10$ and $N_\text{VD}=5$ datasets, respectively. For U-net and MSD we took the same datasets for training and validation (10 for training and 5 for validation), and used all voxels in these datasets for the training process. The NN-FDK networks were trained till convergence and the DNNs were trained for 48 hours. Note that in a few cases we had to retrain the DNNs because of inconsistent results (\ie cases with more information achieving a lower reconstruction accuracy), possibly  because they got stuck in local minima of the loss function.

\begin{figure}[!htp]
  \subfloat[The average and standard deviation of the TSE and SSIM as a function of number of projection angles
  $N_a$ computed over 20 randomly generated phantoms Fourshape family.]{\includegraphics[width=0.48\textwidth]{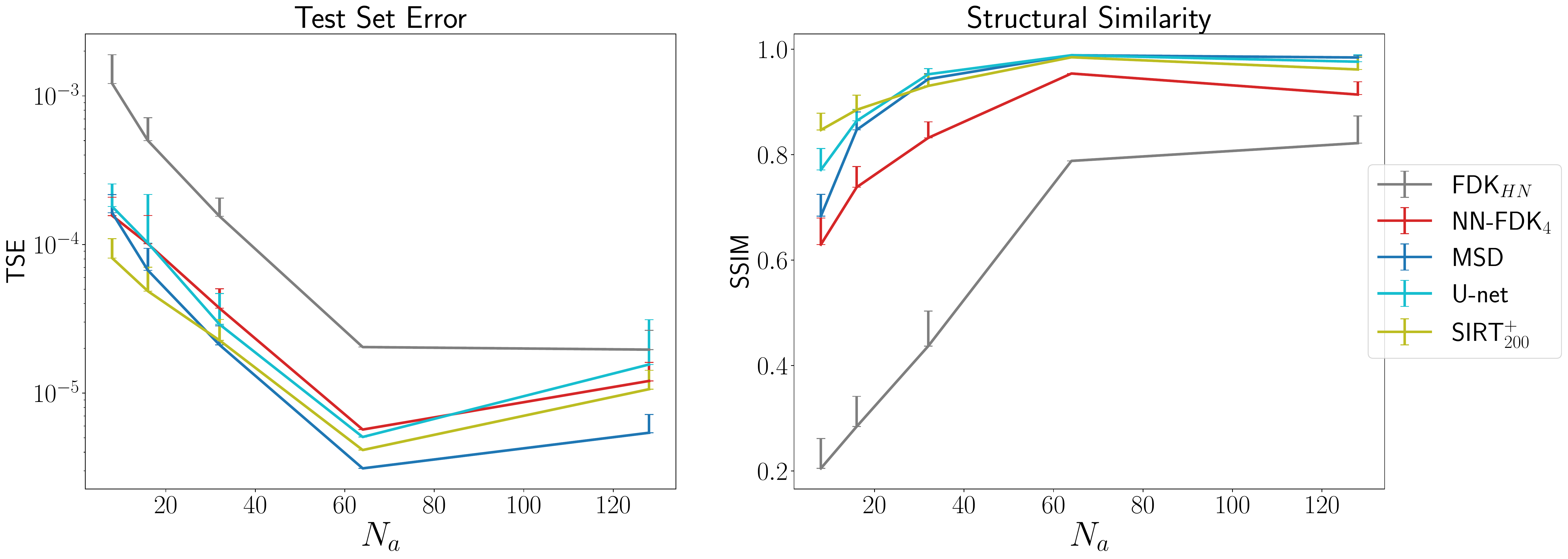}}
  
  \subfloat[The average and standard deviation of the TSE and SSIM as a function of the emitted photon count $I_0$
  computed over 20 randomly generated phantoms of the Fourshape family.]{\includegraphics[width=0.48\textwidth]{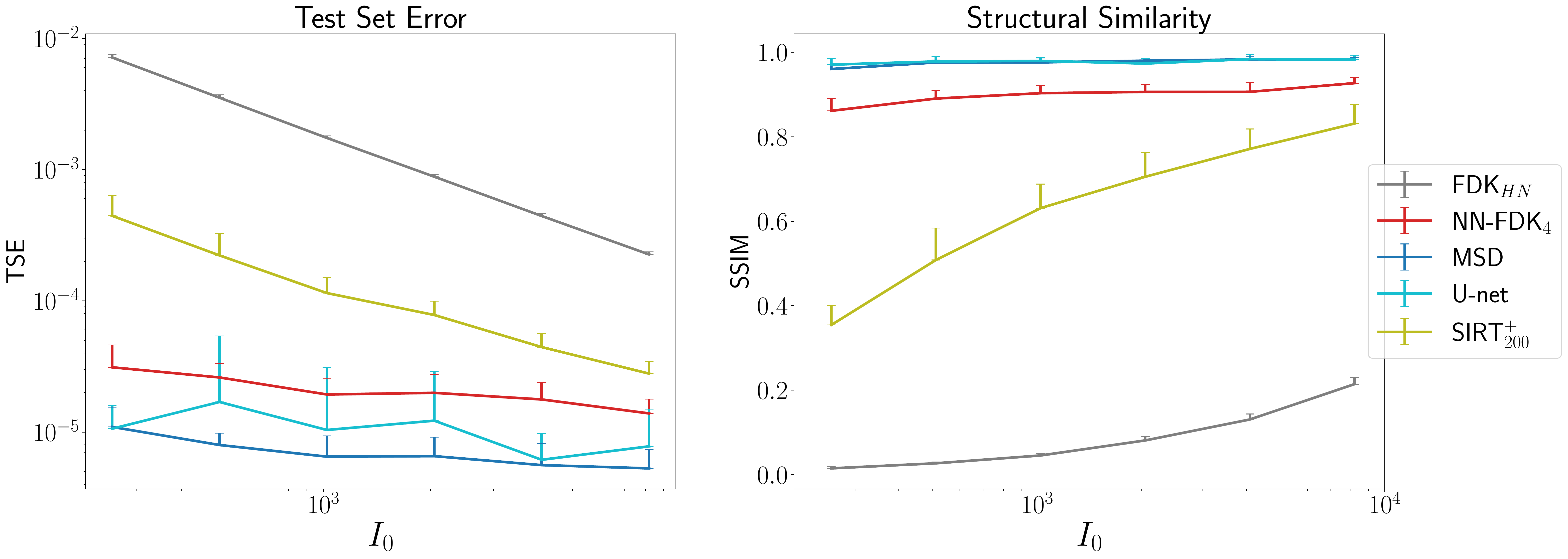}}

  \subfloat[The average and standard deviation of the average TSE and SSIM as a function of the cone angle computed over
  20 randomly generated phantoms of the Defrise
  family.]{\includegraphics[width=.48\textwidth]{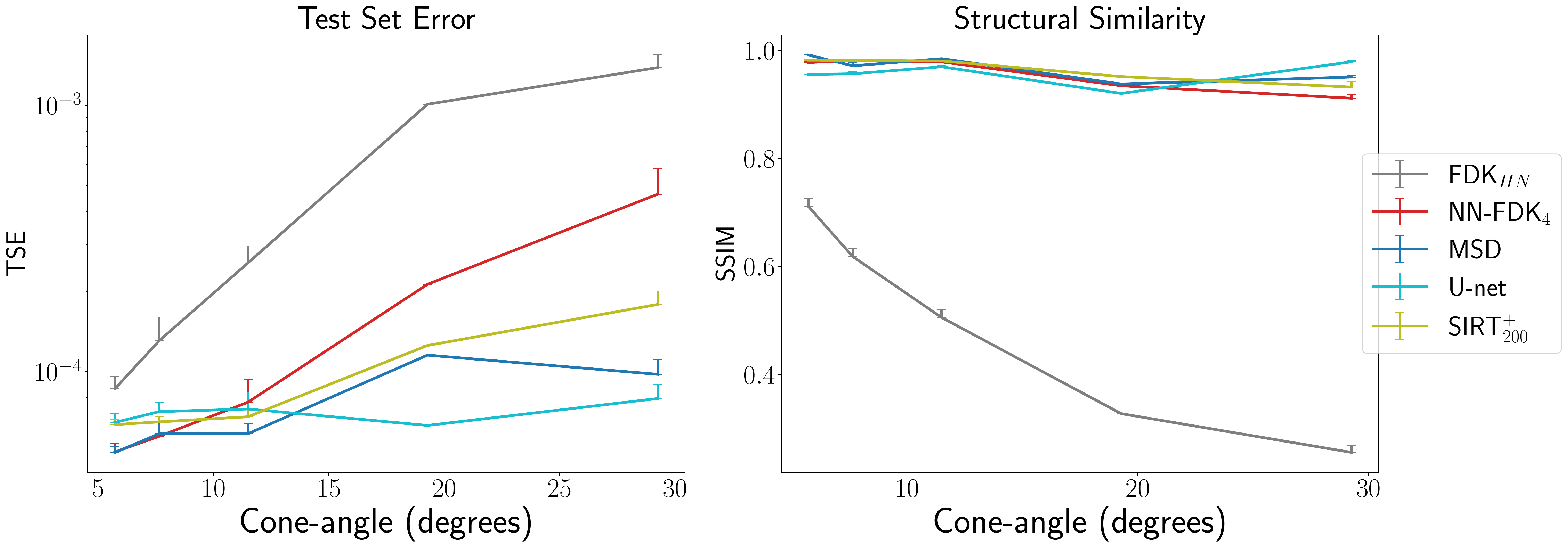}}

\caption{The average and standard deviation of the TSE and SSIM. These results are discussed in \secref{sec:rec_acc_sim}. For each number of projection angles, noise level, cone angle and
  training scenario one specific network is trained and used to evaluate the 20
  reconstruction problems. }
\label{fig:QM_sim}
\end{figure}

\begin{figure}
\centering
    \includegraphics[width=0.48\textwidth]{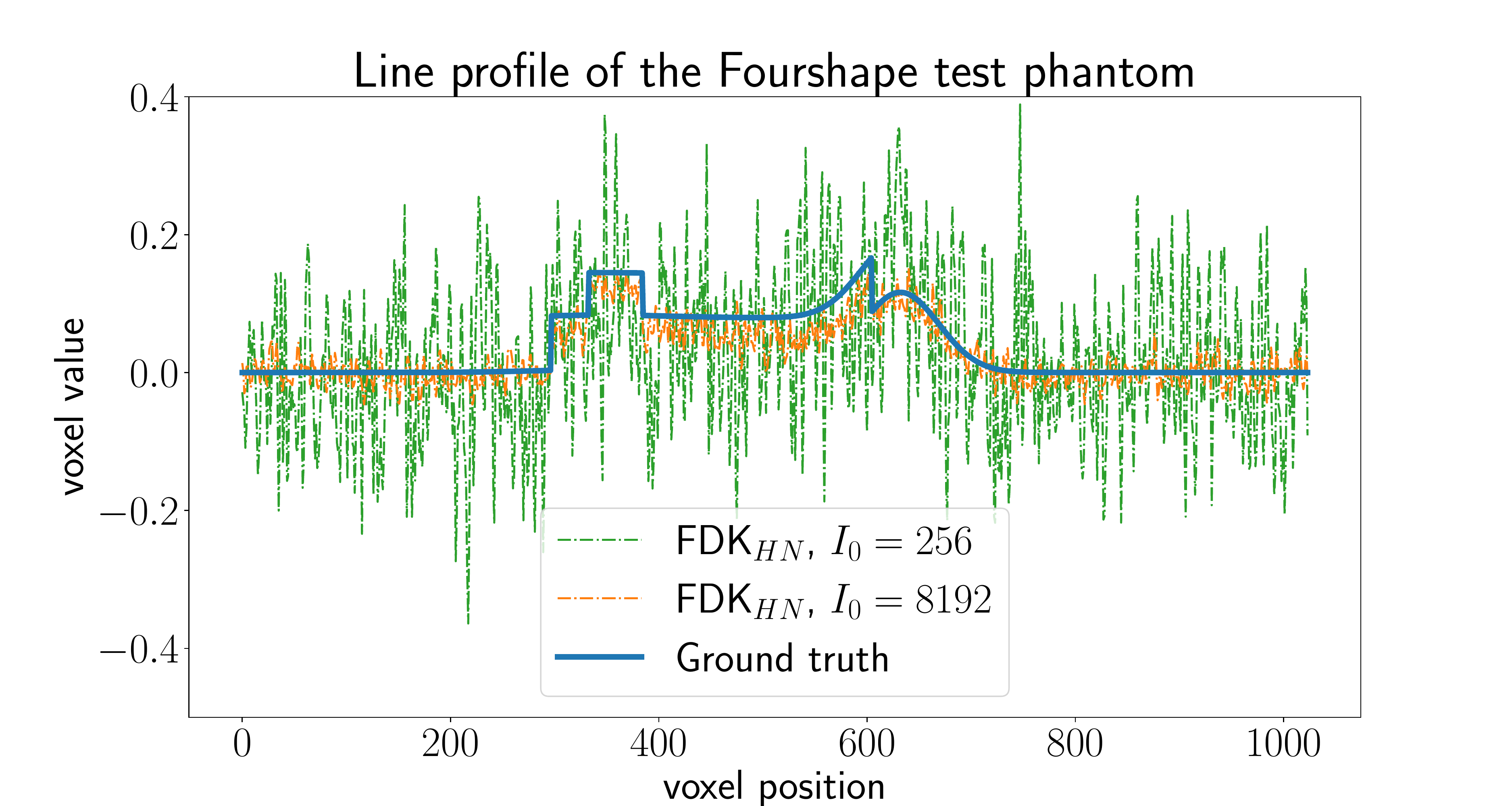}
    \caption{Line profile through the the center of the $z=0$ slice of the Fourshape test phantom. We show the ground truth profile, the profile of the FDK reconstruction with lowest emitted photon count $I_0=256$, and the profile of the FDK reconstruction with the highest emitted photon count $I_0=8196$.}
    \label{fig:range_I0}
\end{figure}

\begin{figure*}
    \includegraphics[width=1\textwidth]{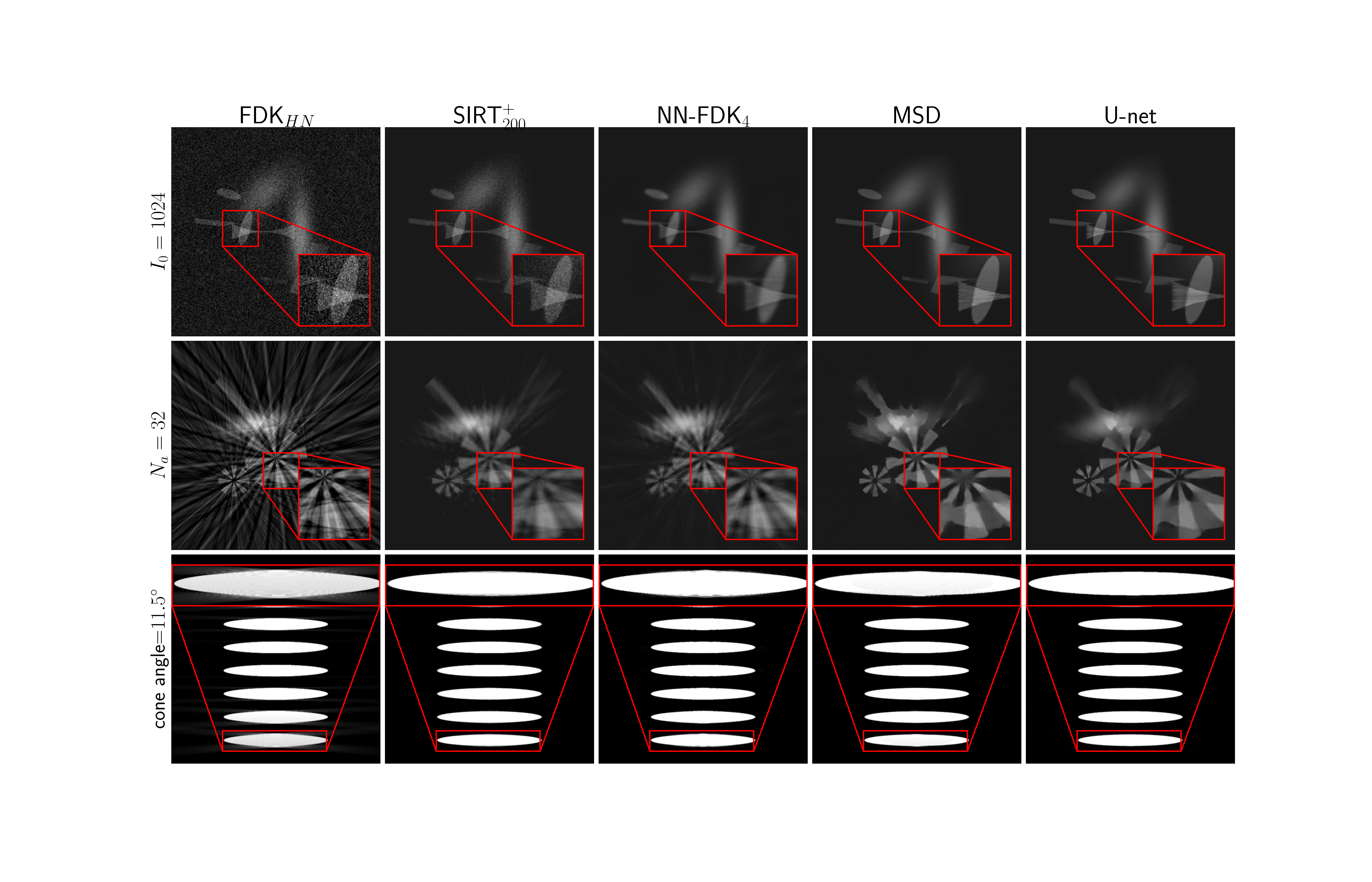}
    \caption{Two-dimensional slices of the reconstructions for the considered reconstruction methods. (Top) Slice $x=0$ of the Fourshape test phantom reconstruction problem with $N_a=360$ projection angles and $I_0=1024$ emitted photon count. (Middle) Slice $z=0$ of the Fourshape test phantom reconstruction problem with $N_a=32$ projection angles. (Bottom) Slice $x=0$ of the Defrise reconstruction problem with $N_a=360$ projection angles and a cone angle of 11.5 degrees.}
    \label{fig:image_grid}
\end{figure*}

In \figref{fig:QM_sim} we show the average and standard deviation of the TSE and the SSIM for the considered cases. We observe that U-net and MSD achieve the most accurate results and that NN-FDK and SIRT$^+$ closely follow. The FDK algorithm is lowest in all categories. Between NN-FDK and SIRT$^+$ we see that NN-FDK performs best for the noisy reconstruction problems and SIRT$^+$ achieves better results for the reconstruction problems without noise. We visualize the noise for the lowest and highest $I_0$ in \figref{fig:range_I0} by showing a line profile through the center of the $z=0$ slice. Here we see that for the noisiest problems the amplitude of the noise can be as high as the maximum value of the phantom.
In \figref{fig:image_grid} we show 2D slices of reconstructions of the test phantoms for the three types of reconstruction problems. In all cases we still observe reconstruction artifacts, but comparing these to the baseline FDK reconstructions, the majority is removed or suppressed. 


\subsection{Reconstruction accuracy for experimental data}
In this section we use the datasets discussed in \secref{sec:exp_data} to assess the reconstruction accuracy on experimental data. In a similar fashion as for the simulated data, we trained a network for the low-dose reconstruction problem and a network for the high-dose reconstruction problem with $N_a=32$ projection angles with the notable exception that U-net and MSD were trained till convergence. The results are presented in \tabref{tab:RD}. 

\begin{figure}[!h]
   \subfloat[FDK$_\text{HN}$]{\includegraphics[width=0.48\textwidth]{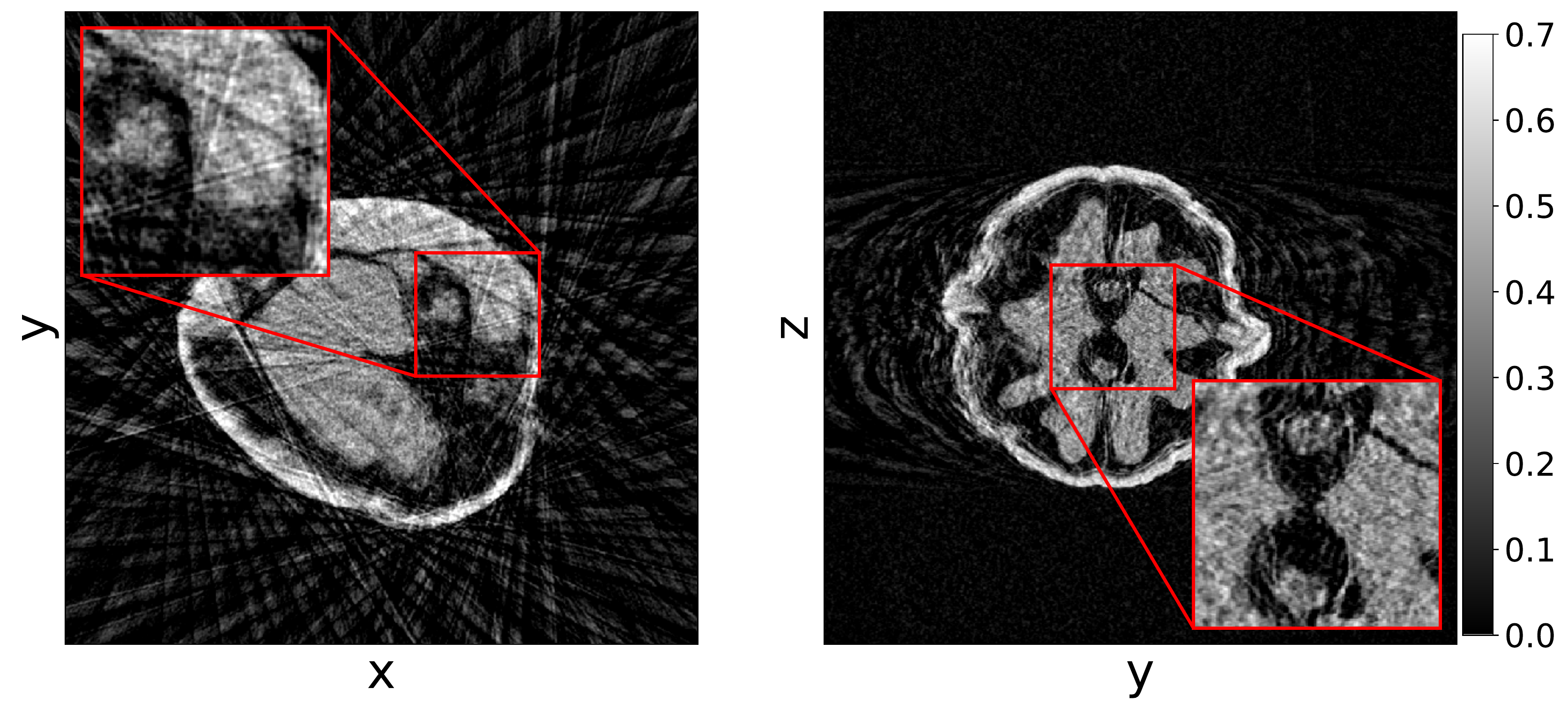}}
   
  \subfloat[SIRT$^+_{200}$
  reconstruction.]{\includegraphics[width=0.48\textwidth]{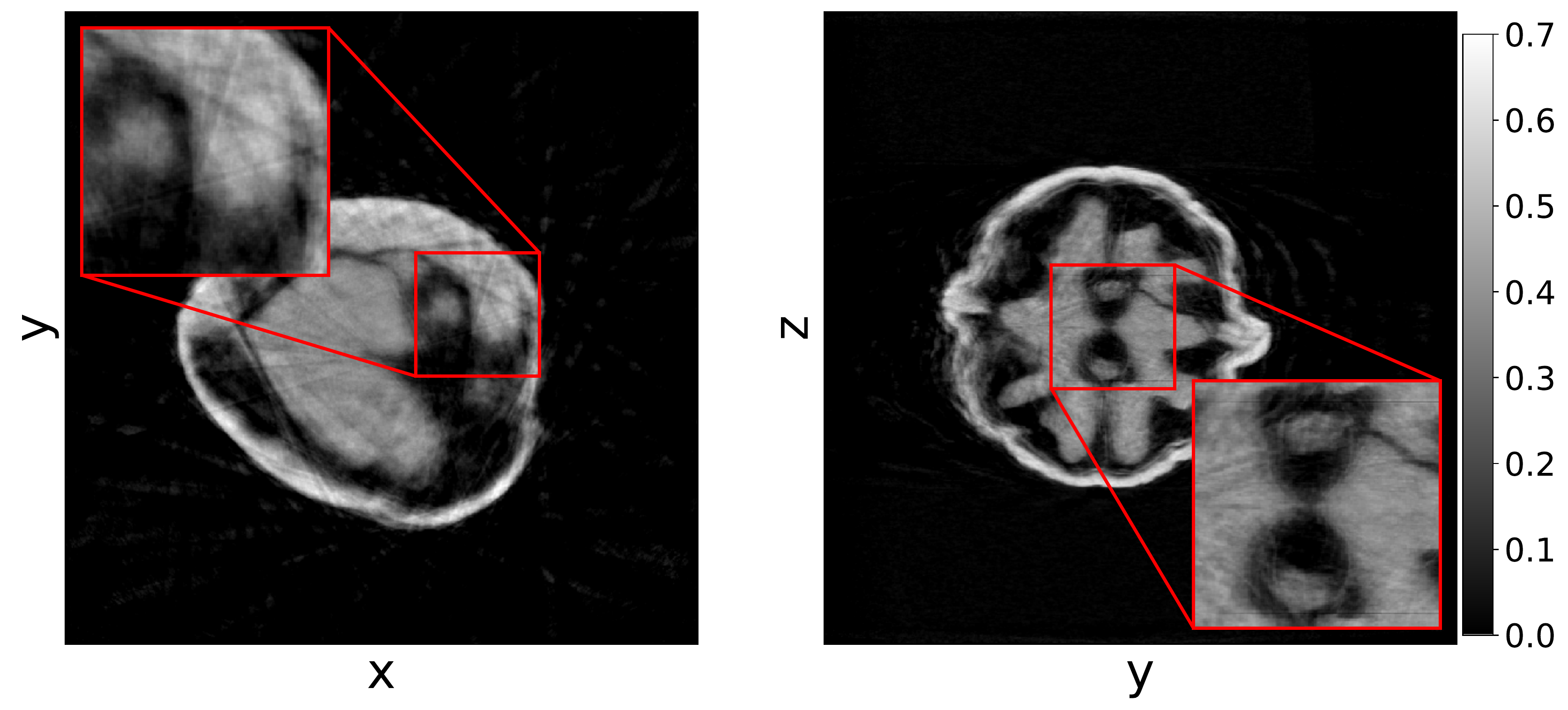}}

  \subfloat[NN-FDK$_4$
  reconstruction.]{\includegraphics[width=0.48\textwidth]{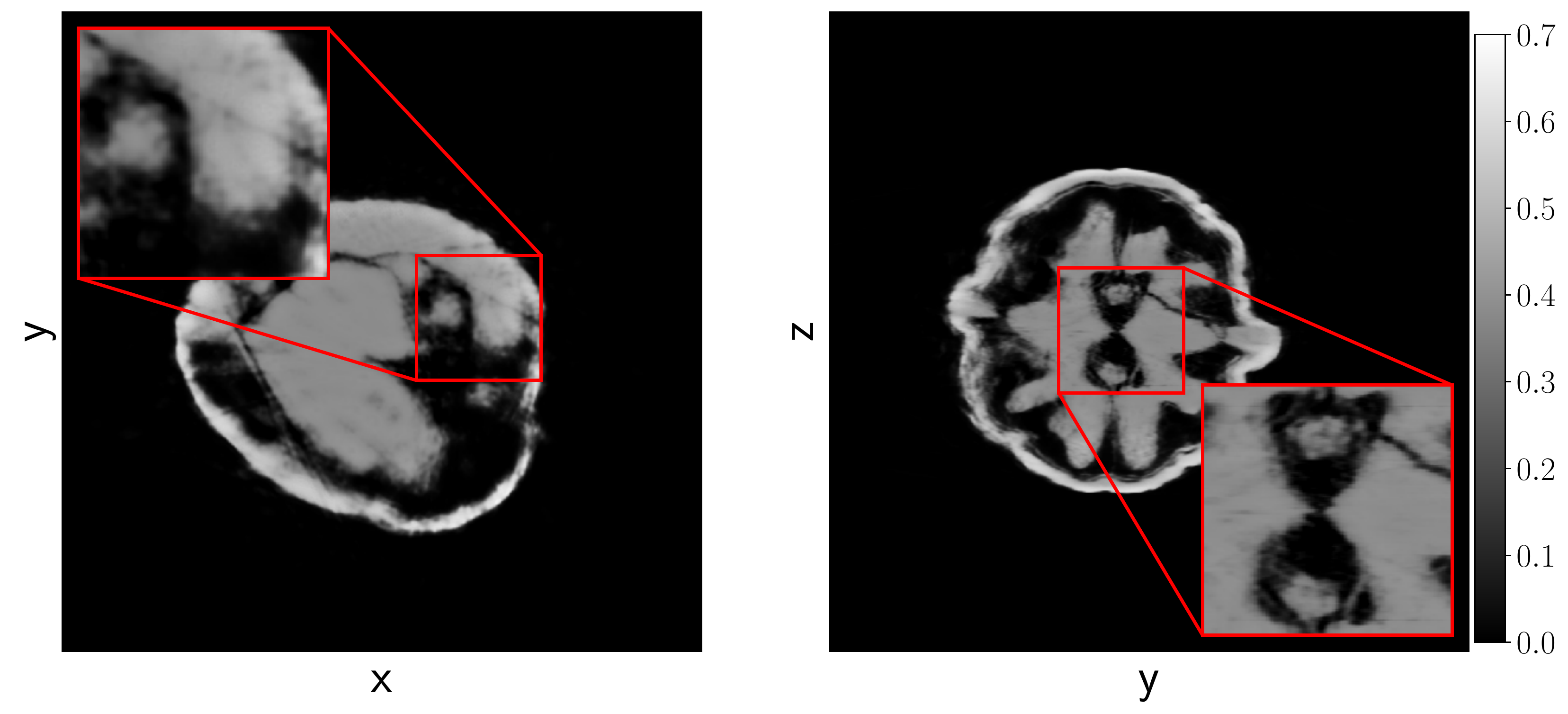}}


 \subfloat[MSD
 reconstruction.]{\includegraphics[width=0.48\textwidth]{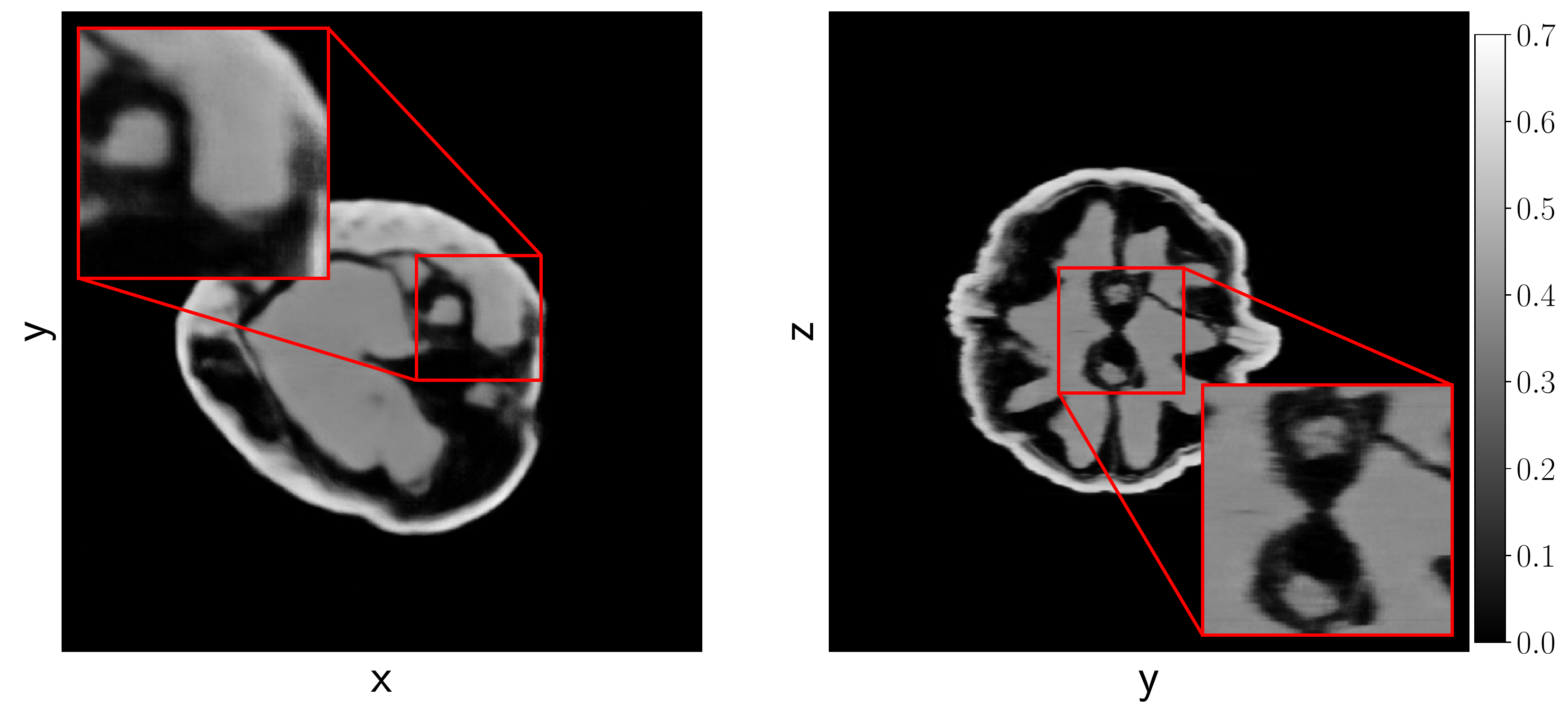}}
\caption{Slices $z=0$ and $x=0$ of several reconstruction methods of the
  high-dose dataset of the $21^{st}$ walnut with 32 projection angles.}
\label{fig:RD_AF16}
\end{figure}

Comparing the results to the simulated data experiments we see that SIRT$^+$ performs worse on the experimental data, even with the additional regularization of early stopping. This is most likely due to the high-dose datasets still containing noise, whereas this is completely absent in the simulated data experiments. These differences are illustrated in  \figref{fig:RD_AF16} where 2D slices of the reconstructions for the high-dose reconstruction problem with $N_a=32$ projection angles are shown. 

\begin{table}[!htp]
  \centering
  \scalebox{0.8}{
  \begin{tabular}{|l||r|r||r|r|}
    \multicolumn{5}{c}{\large{Experimental data}}\\
    \specialrule{.2em}{.05em}{.05em}
    \multicolumn{1}{c}{}& \multicolumn{2}{c}{High-dose, low number} &  \multicolumn{2}{c}{}\\
        \multicolumn{1}{c}{}& \multicolumn{2}{c}{of projection angles} &  \multicolumn{2}{c}{Low-dose}\\
    \hline
    Method           & TSE & SSIM                            & TSE & SSIM\\ 
    \hline
    FDK$_{\text{HN}}$  & 5.54$\pm$3.43e-03 & 0.224$\pm$0.076                      & 1.40$\pm$0.05e-03 & 0.334$\pm$0.104 \\
    SIRT$^+_{200/20}$  & 9.94$\pm$0.15e-04 &  0.603$\pm$0.087                   & 1.92$\pm$0.08e-03 & 0.584$\pm$0.083 \\
    NN-FDK$_4$       & 8.03$\pm$1.39e-04 & 0.946$\pm$0.010                       & 1.14$\pm$0.23e-04 & 0.965$\pm$0.012 \\
    U-net            &\textbf{ 4.10$\pm$1.06e-04} &\textbf{0.964$\pm$0.009}       & 1.02$\pm$0.45e-04 & \textbf{0.980$\pm$0.006}   \\   
    MSD              & 4.23$\pm$0.97e-04 &  \textbf{0.964$\pm$0.009}             & \textbf{7.82$\pm$2.86e-05} &     0.980$\pm$0.007 \\ 
    \hline                         
  \end{tabular}}
\caption{Average and standard deviation of the quantitative measures computed
  over 6 walnut datasets. The high-dose low projection angle reconstruction
  problem has $N_a=32$ projection angles, the low-dose reconstruction problem
  has $N_a=500$ projection angles. The best results per experiment are
  highlighted. }
\label{tab:RD}
\end{table}


\subsection{Segmentation experiment for experimental data}
To assess the performance of the different reconstruction approaches in a segmentation task, we focus here on the segmentation of the shell and kernel of walnuts, based on our experimental CT data. The review \cite{bernard20203d} provides an overview of segmentation problems in walnut imaging, and their relevance. For segmenting the 3D volume after the reconstruction, we used a deterministic segmentation algorithm that combines thresholding, the watershed algorithm and prior knowledge of the scanned objects. Details of this method are discussed in \appref{app:seg_alg}.

For determining the accuracy of the segmentation of an object -- \ie shell, empty space and kernel of the walnut -- we consider three metrics: volume error, mislabeled voxels and the Dice coefficient \cite{dice1945measures}. We define a segmentation $S$ as a reconstruction volume with value 1 if the voxel is in the object (shell, kernel or empty space) and 0 if outside the object. Furthermore we define the norm of a segmentation as the sum:  $\abs{S}=\sum_{i}^{N^3} (S)_i$. Using this notation we can compute the measures in the following manner:
\begin{align}
    V_\text{err} &= \tfrac{\abs{S_\text{rec}} - \abs{S_\text{GS}}}{\abs{S_\text{{GS}}}}, & \text{ML}_\text{err} & = \tfrac{\abs{S_\text{rec}-S_\text{GS}}}{\abs{S_\text{GS}}}, & \text{DC} = \tfrac{2\abs{S_\text{rec}\cap S_\text{GS}}}{\abs{S_\text{rec}} + \abs{S_\text{GS}}}, \label{eq:seg_errs}
\end{align}
with GS denoting the gold standard reconstruction.

In \tabref{tab:seg_errs} we show the results for computing these metrics on the 6 walnuts not considered in the training process. We observe that MSD performs best in segmenting the shell and U-net performs best at segmenting the empty space and kernel and NN-FDK is close to both DNNs and in some cases even better than MSD for segmenting the empty space and kernel. Comparing NN-FDK to standard FDK we observe a significant improvement.

\begin{table}[]
    \centering
    \begin{tabular}{|l|c|c|c|}
    \multicolumn{4}{c}{\normalsize{Segmentation errors}}\\
    \specialrule{.2em}{.05em}{.05em}
    Method & Shell& Empty space & Kernel \\ 
    \hline
    \multicolumn{4}{c}{\normalsize{Volume errors}}\\
\hline
 FDK$_\text{HN}$    &   0.127 $\pm$ 0.078& 0.146 $\pm$ 0.091&    0.128 $\pm$ 0.092\\
 SIRT$^+_{200}$ &   0.082 $\pm$ 0.047& 0.104 $\pm$ 0.078& 0.050 $ \pm$ 0.074 \\
 NN-FDK$_4$  & 0.068 $\pm$ 0.035 & 0.045 $\pm$ 0.035 & 0.029 $\pm$  0.032\\
 U-net     &   0.055 $\pm$ 0.019&         \textbf{0.029 $\pm$ 0.017 }&    \textbf{0.012 $\pm$ 0.016} \\
 MSD       &  \textbf{0.028 $\pm$ 0.010}&         0.059 $\pm$ 0.075&    0.035 $\pm$ 0.050 \\
\hline

    \multicolumn{4}{c}{\normalsize{Mislabeled voxels}}\\
    \hline
 FDK$_\text{HN}$&   0.168 $\pm$ 0.087 &         0.190 $\pm$ 0.98&    0.144 $\pm$ 0.081 \\
 SIRT$^+_{200}$ &   0.133 $\pm$ 0.026&         0.182 $\pm$ 0.118 &    0.101 $\pm$ 0.048\\
 NN-FDK$_{4}$  &   0.103 $\pm$ 0.026&         0.087 $\pm$ 0.023&    0.072 $\pm$ 0.018 \\
 U-net     &   0.092 $\pm$ 0.028&         \textbf{0.073 $\pm$ 0.024}&    \textbf{0.059 $\pm$ 0.019}\\
 MSD       &   \textbf{0.086 $\pm$ 0.038}&         0.116 $\pm$ 0.094&    0.061 $\pm$ 0.039\\
    \hline
    \multicolumn{4}{c}{\normalsize{Dice coefficient}}\\
    \hline
 FDK$_\text{HN}$    &   0.922 $\pm$ 0.036&         0.895 $\pm$ 0.061&    0.934 $\pm$ 0.033\\
 SIRT$^+_{200}$ &   0.934 $\pm$ 0.016&         0.908 $\pm$ 0.061&    0.947 $\pm$ 0.028\\
 NN-FDK$_4$  &   0.951 $\pm$ 0.012&         0.955 $\pm$ 0.013&    0.964 $\pm$ 0.008 \\
 U-net     &   0.955 $\pm$ 0.013&        \textbf{0.963 $\pm$ 0.012}&    \textbf{0.971 $\pm$ 0.010}\\
 MSD       &   \textbf{0.957 $\pm$ 0.018}&         0.939 $\pm$ 0.055&    0.971 $\pm$ 0.018 \\
    \hline
    \end{tabular}
    \caption{The average and standard deviation of the three metrics computed over the 6 low-dose walnut datasets with $N_a=500$ projection angles. The metrics are computed using \eqref{eq:seg_errs}. The best results are highlighted.}
    \label{tab:seg_errs}
\end{table}

    

\subsection{Data requirements}\label{sec:data_req}
To test the influence of the amount of training data on the reconstruction quality we performed an experiment
with three different training scenarios:
\begin{itemize}
\item \textbf{Scenario 1}. One dataset available. Here we take the
  training and validation data from the same dataset.
\item \textbf{Scenario 2}. Two datasets available. Here we take the
  training and validation data from the separate datasets.
\item \textbf{Scenario 3}. Fifteen datasets available. Again the
  training and validation data are picked from separate datasets, but now the
  training and validation pairs come from several datasets, specifically 10 training datasets
  ($N_{\text{TD}}=10$) and 5 validation datasets ($N_{\text{VD}}=5$). This is
  the scenario used in the previous experiments.
\end{itemize}
We fix the number of voxels used for training and validation at
$N_{\text{T}}=10^6$ and $N_{\text{V}}=10^6$ for all scenarios. For comparison we
trained a U-net and a MSD network with the same training scenarios, with the
exception that all voxels from the datasets are used. For training scenario 1
the slices are divided into a training and a validation set. More
specifically, every fourth slice is used for validation. 

We performed this experiment for two simulated data problems, a high noise level
(emitted photon count $I_0=256$) and a large cone angle (29.3 degrees), and the two experimental
data problems. For the sake of brevity we
show only the results for the high noise simulated data reconstruction
problem (\tabref{tab:sim_scens}) and the high noise experimental data
reconstruction problem (\tabref{tab:exp_scens}). The results for the other
reconstruction problems are given in \appref{sec:scens}. Comparing quantitative
measures between the different scenarios we see that the reconstruction accuracy
improves as more data is used for the simulated data experiment, but remains
about the same for the experimental data experiment. This can be explained by
the variation in the objects used in the reconstruction problems. Recall that
the Fourshape phantom family has a large variety in its phantoms, \ie three
instances of four randomly generated objects, and the variety within the walnut
datasets is small, \ie similar shapes, sizes and structures. This indicates that
if objects are similar, one training dataset may already be sufficient to train
networks that achieve a high reconstruction accuracy.

Note that although the training scenarios for NN-FDK and the DNNs use the same
number of datasets, the number of voxels considered for training the NN-FDK
network is constant over all three scenarios and is several orders of magnitude
lower than the number of voxels considered for training the DNNs. This opens up
future possibilities for reducing the training data requirements to only need a
high quality reconstruction of a certain region of interest.

\begin{table}[!h]
    \begin{tabular}{|l|r|r|r|}
    \multicolumn{4}{c}{\normalsize{Simulated data, high noise}}\\
    \specialrule{.2em}{.05em}{.05em}
    \multicolumn{4}{c}{TSE}\\
    \hline
    {Method}&  {1 dataset} &{2 datasets} & {15 datasets} \\
    \hline
    \hline
    NN-FDK$_{4}$ & 4.97$\pm$4.68e-05 & 4.19$\pm$3.60e-05  & 2.51$\pm$1.14e-05  \\
    \hline
    U-net & 1.06$\pm$1.36e-05&   2.45$\pm$2.87e-05 &   8.06$\pm$3.63e-06 \\
    \hline
    MSD  & 1.12$\pm$0.41e-05  &  1.12$\pm$0.40e-05 & \textbf{7.94$\pm$3.16e-06}\\
    \hline
    \multicolumn{4}{c}{SSIM}\\
    \hline
    NN-FDK$_{4}$ & 0.831$\pm$0.065  & 0.844$\pm$0.065& 0.884$\pm$0.030\\
    \hline
    U-net &   0.884$\pm$0.075 &   0.932$\pm$0.050 & \textbf{0.979$\pm$0.009}\\
    \hline
    MSD& 0.961$\pm$0.013 &  0.962$\pm$0.013  &   0.974$\pm$0.008\\
    \hline
  \end{tabular}
  \caption{Average and standard deviation of the quantitative measures computed
    over 20 Fourshape phantoms for varying training scenarios. The
    reconstruction problems have an emitted photon count of $I_0=256$ and
    $N_a=360$ projection angles. The best results are highlighted.}
\label{tab:sim_scens}
\end{table}
\begin{table}[!h]
  \centering
  \begin{tabular}{|l|r|r|r|}
    \multicolumn{4}{c}{\normalsize{Experimental data, low-dose}}\\
    \specialrule{.2em}{.05em}{.05em}
    \multicolumn{4}{c}{TSE}\\
    \hline
    {Method}&  {1 dataset} &{ 2 datasets} & {15 datasets} \\
    \hline
    \hline
    NN-FDK$_{4}$ &  1.16$\pm$0.25e-04 & 1.23$\pm$0.25e-04 & 1.14$\pm$0.23e-04\\ 
    \hline
    U-net & 1.27$\pm$0.38e-04  & 1.23$\pm$0.35e-04   & 1.02$\pm$0.45e-04\\ 
    \hline
    MSD & 1.28$\pm$0.41e-04 & 1.16$\pm$0.35e-04 & \textbf{7.82$\pm$2.86e-05}   \\
    \hline
    \multicolumn{4}{c}{SSIM}\\
    \hline
    NN-FDK$_{4}$ & 0.973$\pm$0.009  & 0.968$\pm$0.011  & 0.965$\pm$0.012\\
    \hline
    U-net &    0.979$\pm$0.008 & 0.978$\pm$0.008 &  \textbf{0.980$\pm$0.006}\\
    \hline
    MSD&         0.979$\pm$0.008 &       0.979$\pm$0.008  & 0.980$\pm$0.007 \\     
    \hline
  \end{tabular}
  \caption{Average and standard deviation of the quantitative measures computed
    over 6 walnuts for varying training scenarios. The datasets are low-dose and
    have $N_a=500$ projection angles. The best results are highlighted.}
  \label{tab:exp_scens}
\end{table}

\section{Summary \& Conclusion}\label{sec:conclusions}
We have proposed the Neural Network FDK (NN-FDK) algorithm, a reconstruction
algorithm for the circular cone-beam (CCB) Computed Tomography (CT) geometry
with a machine learning component. The machine learning component of the
algorithm is designed to learn a set of FDK filters and to combine the FDK
reconstructions done with these filters. This leads to a computationally
efficient reconstruction algorithm, since one only needs to compute and combine
the FDK reconstructions for this learned set of filters. Due to parametrization
of the learned filters, the NN-FDK network has a low number of trainable
parameters (${<}100$) and can be trained efficiently with the
Levenberg-Marquardt algorithm with approximate quadratic convergence rate.

We compared the NN-FDK algorithm to SIRT with a nonnegativity constraint
(SIRT$^+$), the standard FDK algorithm and two deep neural networks (DNNs),
namely a 2D U-net and a 2D MSD network applied in a slice-by-slice fashion to a
3D volume. We have shown that the NN-FDK algorithm has the lowest reconstruction
time after the standard FDK algorithm. We have also shown that the NN-FDK
algorithm achieves a reconstruction accuracy that is similar to that of SIRT$^+$
for simulated data and a higher accuracy than that of SIRT$^+$ for experimental
data. The DNNs achieved the highest reconstruction accuracy, but training those
networks took between 2 days (1 training and validation dataset) and 2 weeks (15
training and validation datasets), whereas all the NN-FDK networks were trained
within 1 minute.

To conclude, the NN-FDK algorithm is a computationally efficient reconstruction
algorithm that can reconstruct CCB CT reconstruction problems with high noise,
low projection angles or large cone angles accurately. The training process is
efficient and requires a low amount of training data, making it suitable for
application to a broad spectrum of large scale (up to $4096\times4096\times4096$) reconstruction problems. Specifically, the NN-FDK algorithm can be used improve image quality in high throughput CT scanning settings, where FDK is currently used to keep pace with the acquisition speed using readily available computational resources. 
\section*{Acknowledgements}
The authors acknowledge financial support from the Netherlands Organisation for
Scientific Research (NWO), project numbers 639.073.506 and 016.Veni.192.235. We
acknowledge XRE NV for their role in the FleX-ray collaboration. We thank Sophia
Bethany Coban for her support in acquiring the experimental data.

\appendices
  
\section{Implementation}\label{sec:implementation}
\subsection{Data generation}\label{sec:data_gen}
For our simulated data experiments we take $N=1024$, which means that
reconstructions and reference images are defined on a $1024^3$ equidistant voxel
grid, and the projection data on a $1024^2$ equidistant detector grid per
projection angle. However, to avoid using the same operator for
reconstructions as for the data generation we generate the input data at a
higher resolution. More specifically, we generate a phantom at $N=1536$, forward
project this phantom to the data space with size $N_a\times 1536^2$ and apply a
bilinear interpolation per projection angle to arrive at a $1024^2$ detector
grid, resulting in input data with the desired resolution $N_a\times 1024^2$. We
set the source radius to 10 times the physical size of the phantom, resulting in
a cone angle of $5.7$ degrees. To generate noise we compute a noise free photon
count $I$ from clean projection data $\y_c$ and use that to generate a Poisson
distributed photon count from which we compute $\y$:
\begin{align}
I &= I_0e^{-\y_c},  &I_{\textnormal{noise}}\sim \text{Pois}(I),& &\mathbf{y}&= -\log \brac{\frac{I_{\textnormal{noise}}}{I_0}},
\end{align}
with $I_0$ the emitted photon count. Higher $I_0$ implies a higher dose and
therefore less noise in the data.

\subsection{Deep neural networks}\label{sec:DNNs}
\subsubsection{Application strategy} We train 2D DNNs to remove artifacts from
2D slices of an FDK reconstruction. We train one network that handles all slices
in the reconstructions.

\subsubsection{Training DNNs} We train the DNNs with ADAM \cite{kingma2014adam}
and stop training after 48 hours of training on a Nvidia GeForce GTX 1080Ti GPU, the network with the lowest validation set error during this training process will be used for the reconstructions.

\subsubsection{U-net and MSD network structures} For U-net we will take four up
and down layers as presented in \cite{ronneberger2015u}. For the MSD networks we
take 100 layers with one input and one output layer and the dilations as
suggested in \cite{pelt2018mixed}.

\subsection{Code-base}
We implemented the NN-FDK framework using Python 3.6.5 and Numpy 1.14.5
\cite{walt2011numpy}. For the parameter learning we used the Levenberg-Marquardt
algorithm implementation from \cite{pelt2013fast}. The reconstruction algorithm
is implemented using ODL \cite{odl}, the ASTRA-toolbox \cite{van2016fast},
PyFFTW \cite{frigo2005design} and the exponential binning framework for filters
from \cite{lagerwerf2020automated}. For performance reasons the simulated
phantoms are generated through C++ using Cython \cite{behnel2010cython}.

For the evaluation of U-nets we took the PyTorch \cite{NEURIPS2019_9015}
implementation used in \cite{hendriksen2019fly}. The MSD-nets are implemented
using the package published with \cite{pelt2018improving}.

All the code related to this paper can be found on Github \cite{nnfdk}.

\subsection{Segmentation algorithm}\label{app:seg_alg}
This algorithm consists of several steps:
\begin{enumerate}
    \item Apply a Gaussian filter to the reconstruction.
    \item Compute a histogram of the filtered reconstruction and determine the peaks relating to the background, kernel and shell.
    \item Determine the shell and kernel segmentations using a threshold based on the found peaks.
    \item Apply the watershed algorithm on the shell segmentation. This gives the total volume inside the walnut.
    \item Remove the kernel from the total volume inside the walnut to attain the empty space segmentation.
\end{enumerate}
Further details about this implementation can be found on our Github \cite{nnfdk}.
\section{Levenberg-Marquardt algorithm}\label{sec:LMA}
Given the learning problem \eqref{eq:MLP_MP}, the update rule for the Levenberg-Marquardt algorithm (LMA)
(\cite{levenberg1944method, marquardt1963algorithm}) is given by:
\begin{align}
  \mathbf{\theta}^{i+1} = \mathbf{\theta}^{i} + \mathbf{t}^i,
\end{align}
with $\mathbf{t}^i$ the update vector. This is computed by solving the following
equation for $\mathbf{t}^i$
\begin{align}
  \brac{J_i^TJ_i +\lambda_iI}\mathbf{t}^i =-\pderiv{\mathcal{L}}{\mathbf{\theta}}(\mathbf{\theta}^i, T)=-J^T_i\sum^{N_{\text{T}}}_{j=1}\brac{O_j-\mathsf{N}_{\theta}(Z_j)}\label{eq:LMA}
\end{align}
where $\lambda_i>0$ is the step parameter and $J_i$ the $m\times n$ Jacobian
matrix of $\mathsf{N}_{\theta^i}(\mathbf{Z})$ with respect to $\theta^i$, with
$\mathbf{Z}$ the vector containing all inputs from the training set $T$. We can
solve \eqref{eq:LMA} using a Cholesky decomposition.\footnote{$J_i^TJ_i$ is
  positive semi-definite and $\lambda_i>0$, therefore the left hand side of
  \eqref{eq:LMA} is positive definite.}

To ensure convergence, only updates that improve the training error are accepted,
\ie if the following is true:
\begin{align}
  \mathcal{L}(\theta^i, T)>\mathcal{L}(\theta^{i}+\mathbf{t}^i,T),\label{eq:LMA_update}
\end{align}
If this is not the case we change the step parameter $\lambda_i$ to $
a\lambda_i$ with $a>1$ and compute a new update vector $\mathbf{t}^i$. When an update is
accepted we change the step parameter to $\lambda_{i+1}=\lambda_i/a$.

We use two stopping criteria for the LMA. Firstly, we stop if we cannot find a
suitable $\theta^{i+1}$, using several indicators for this:
\begin{itemize}
\item The norm of the gradient $\pderiv{\mathcal{L}}{\theta}(\theta^i)$ is too small
\item The step size $\lambda_i$ is too big
\item After $N_{\text{up}}$ rejected updates.
\end{itemize}
The second stopping criterion checks whether the parameters $\theta^i$ improve the
validation set error. More specifically, we terminate the LMA when the
validation set error has not improved for $N_{\text{val}}$ iterations.

In \alref{alg:LMA} the LMA is summarized. The random initialization is done with
the Nguyen-Widrow initialization method \cite{nguyen1990truck}. For our
experiments we take $N_{\text{up}}=100$, $\lambda_0=10^5$, $a=10$ and
$N_{\text{val}}=100$.

\begin{algorithm}
  \caption{Levenberg-Marquardt algorithm}
  \begin{algorithmic}[1]
    \STATE{Compute random initialization $\theta^0$ using
      \cite{nguyen1990truck}}
    \REPEAT
    \STATE{Compute $\mathbf{t}^i$ until we
      accept an update $\theta^{i+1}$.}
    \UNTIL{$N_{\text{up}}$ updates were rejected \textbf{or}\\
      $\mathcal{L}(\theta^i, V)$ did not improve $N_{\text{val}}$ times \textbf{or} \\
      $\norm{\pderiv{\mathcal{L}}{\theta}(\theta^{i+1})}$ is
      too small \textbf{or} $\lambda_{i+1}$ is too big.}
    \STATE{Set
      $\theta^{\star}$ equal to the $\theta^i$ with the lowest
      validation error.}
  \end{algorithmic}\label{alg:LMA}
\end{algorithm}

\section{Results data requirement experiment}\label{sec:scens}
\FloatBarrier
\begin{table}[!h]
  \centering
  \begin{tabular}{|l|r|r|r|}
    \multicolumn{4}{c}{\normalsize{Simulated data, large cone angle}}\\
    \specialrule{.2em}{.05em}{.05em}
    \multicolumn{4}{c}{TSE}\\
    \hline
    {Method}&  {1 dataset} &{ 2 datasets} & {15 datasets} \\
    \hline
    \hline
    NN-FDK$_{4}$  &  6.47$\pm$1.19e-04 & 4.70$\pm$1.16e-04 & 4.82$\pm$1.13e-04 \\ 
    \hline
    U-net        &  1.04$\pm$0.27e-04  &   1.02$\pm$0.17e-04 & 8.23$\pm$0.85e-05 \\
    \hline
    MSD          &  2.44$\pm$1.43e-04& 1.53$\pm$0.17e-04  &   \textbf{6.52$\pm$0.43e-05}  \\
    \hline
    \multicolumn{4}{c}{SSIM}\\
    \hline
    NN-FDK$_{4}$ &  0.825$\pm$0.018  &  0.904$\pm$0.011  & 0.910$\pm$0.007 \\
    \hline
    U-net &      \textbf{0.974$\pm$0.015} &   0.971$\pm$0.021 &  0.973$\pm$0.010 \\
    \hline
    MSD&        0.954$\pm$0.006    &   0.937$\pm$0.004     &  0.966$\pm$0.002\\     
    \hline
  \end{tabular}
  \caption{Average and standard deviation of the quantitative measures
      computed over 20 different Defrise phantoms for varying training
      scenarios. The reconstruction problems have a cone angle of 29.2 degrees
      and $N_a=360$ projection angles. The best results are highlighted.}
\end{table}
\begin{table}[!h]
  \centering
  \begin{tabular}{|l|r|r|r|}
    \multicolumn{4}{c}{\normalsize{Experimental data, high-dose, 32 projection angles}}\\
    \specialrule{.2em}{.05em}{.05em}
    \multicolumn{4}{c}{TSE}\\
    \hline
    {Method}&  {1 dataset} &{ 2 datasets} & {15 datasets} \\
    \hline
    \hline
    NN-FDK$_{4}$  &   8.14$\pm$1.45e-04&  8.68$\pm$1.43e-04 &  8.03$\pm$1.39e-04 \\ 
    \hline
    U-net        &  7.56$\pm$1.52e-04&  6.85$\pm$1.56e-04 &\textbf{4.10$\pm$1.06e-04} \\
    \hline
    MSD          &  7.82$\pm$0.41e-04  &  6.51$\pm$0.35e-04 &  4.23$\pm$0.97e-04  \\
    \hline
    \multicolumn{4}{c}{SSIM}\\
    \hline
    NN-FDK$_{4}$ &  0.950$\pm$0.010&  0.948$\pm$0.010 & 0.946$\pm$0.011 \\
    \hline
    U-net      &   0.955$\pm$0.011   &  0.930$\pm$0.023  & \textbf{0.964$\pm$0.009} \\
    \hline
    MSD       &     0.955$\pm$0.010     &  0.947$\pm$0.014 & \textbf{0.964$\pm$0.009} \\     
    \hline
  \end{tabular}
  \caption{Average and standard deviation of the quantitative measures computed over
      the 6 datasets for varying training scenarios. These are the
      high-dose datasets from \cite{walnuts} with $N_a=32$ projection angles. The best
      results are highlighted.}
\end{table}
\FloatBarrier

\begin{IEEEbiography}[{\includegraphics[width=1in,height=1.25in,clip,keepaspectratio]{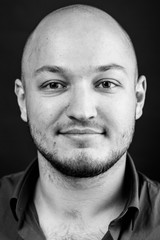}}]{Marinus
   J. Lagerwerf}
 received the M.Sc. degree in applied mathematics from the University of
 Twente, The Netherlands in 2015. He is currently pursuing a Ph.D. degree with
 the Computational Imaging group at CWI, the national research institute for
 mathematics and computer science in Amsterdam, The Netherlands, focusing on
 tomographic reconstruction algorithms.
\end{IEEEbiography}
\begin{IEEEbiography}[{\includegraphics[width=1in,height=1.25in,clip,keepaspectratio]{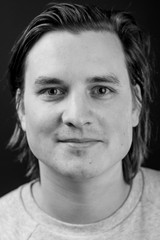}}]{Dani\"el Pelt }
  Dani\"el M. Pelt received the M.Sc. degree in mathematics from the University
  of Utrecht, Utrecht, The Netherlands in 2010, and the Ph.D. degree at Leiden
  University, Leiden, The Netherlands, in 2016. As a Postdoctoral Researcher, he
  was at the Lawrence Berkeley National Laboratory (2016 - 2017), focusing on
  developing machine learning algorithms for imaging problems. He is currently a
  Postdoctoral Researcher with the CWI. His main research interest is developing
  machine learning algorithms for imaging problems, including tomographic
  imaging.
\end{IEEEbiography}
\begin{IEEEbiography}[{\includegraphics[width=1in,height=1.25in,clip,keepaspectratio]{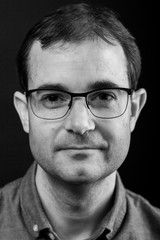}}]{Willem
   Jan Palenstijn}
 Willem Jan Palenstijn received the M.Sc. and Ph.D. degrees in mathematics from
 Universiteit Leiden, in 2004 and 2014, respectively. He has been a Research
 Assistant and Post-Doctoral Researcher on tomographic image reconstruction and
 GPU computing at the University of Antwerp and at CWI, the national research
 institute for mathematics and computer science in Amsterdam, The Netherlands.
 Currently he is a Scientific Software Developer at CWI, and one of the lead 
 developers of the ASTRA Tomography Toolbox.
\end{IEEEbiography}
\begin{IEEEbiography}[{\includegraphics[width=1in,height=1.25in,clip,keepaspectratio]{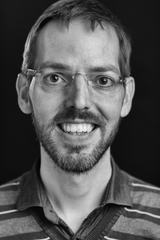}}]{K.
   Joost Batenburg} heads the Computational Imaging group at CWI, the national
 research center for mathematics and computer science in The Netherlands. Most
 of his research concerns various mathematical and computational aspects of
 tomography. He is responsible for the FleX-Ray lab, where a custom-designed CT
 system is linked to advanced data processing and reconstruction algorithms.
\end{IEEEbiography}

\bibliographystyle{IEEEtran}
\bibliography{bibliography}

\begin{thebibliography}{10}
\providecommand{\url}[1]{#1}
\csname url@samestyle\endcsname
\providecommand{\newblock}{\relax}
\providecommand{\bibinfo}[2]{#2}
\providecommand{\BIBentrySTDinterwordspacing}{\spaceskip=0pt\relax}
\providecommand{\BIBentryALTinterwordstretchfactor}{4}
\providecommand{\BIBentryALTinterwordspacing}{\spaceskip=\fontdimen2\font plus
\BIBentryALTinterwordstretchfactor\fontdimen3\font minus
  \fontdimen4\font\relax}
\providecommand{\BIBforeignlanguage}[2]{{%
\expandafter\ifx\csname l@#1\endcsname\relax
\typeout{** WARNING: IEEEtran.bst: No hyphenation pattern has been}%
\typeout{** loaded for the language `#1'. Using the pattern for}%
\typeout{** the default language instead.}%
\else
\language=\csname l@#1\endcsname
\fi
#2}}
\providecommand{\BIBdecl}{\relax}
\BIBdecl

\bibitem{giudiceandrea2011high}
F.~Giudiceandrea, E.~Ursella, and E.~Vicario, ``A high speed {CT} scanner for
  the sawmill industry,'' in \emph{Proceedings of the 17th international non
  destructive testing and evaluation of wood symposium}.\hskip 1em plus 0.5em
  minus 0.4em\relax University of West Hungary Sopron, Hungary, 2011, pp.
  14--16.

\bibitem{dierick2014recent}
M.~Dierick, D.~Van~Loo, B.~Masschaele, J.~Van~den Bulcke, J.~Van~Acker,
  V.~Cnudde, and L.~Van~Hoorebeke, ``Recent {micro-CT} scanner developments at
  {UGCT},'' \emph{Nuclear Instruments and Methods in Physics Research Section
  B: Beam Interactions with Materials and Atoms}, vol. 324, pp. 35--40, 2014.

\bibitem{bultreys2016fast}
T.~Bultreys, M.~A. Boone, M.~N. Boone, T.~De~Schryver, B.~Masschaele,
  L.~Van~Hoorebeke, and V.~Cnudde, ``Fast laboratory-based micro-computed
  tomography for pore-scale research: illustrative experiments and perspectives
  on the future,'' \emph{Advances in water resources}, vol.~95, pp. 341--351,
  2016.

\bibitem{ford2002cone}
E.~Ford, J.~Chang, K.~Mueller, K.~Sidhu, D.~Todor, G.~Mageras, E.~Yorke,
  C.~Ling, and H.~Amols, ``Cone-beam {CT} with megavoltage beams and an
  amorphous silicon electronic portal imaging device: Potential for
  verification of radiotherapy of lung cancer,'' \emph{Medical physics},
  vol.~29, no.~12, pp. 2913--2924, 2002.

\bibitem{galicia2017clinical}
J.~C. Galicia, J.~Kawilarang, and P.~Z. Tawil, ``Clinical endodontic
  applications of cone beam-computed tomography in modern dental practice,''
  \emph{Open Journal of Stomatology}, vol.~7, no.~07, p. 314, 2017.

\bibitem{xre2019unitom}
{TESCAN}, ``{TESCAN} uni{TOM} xl, {M}odular and versatile high resolution {3D}
  {X-}ray imaging,''
  \url{https://www.tescan.com/product/micro-ct-for-materials-science-tescan-unitom-xl/},
  [Accessed: 20-May-2020].

\bibitem{xre2019dynatom}
------, ``{TESCAN} dyna{TOM}, {H}igh temporal resolution {4D} {X-}ray
  imaging,''
  \url{https://www.tescan.com/product/micro-ct-for-materials-science-tescan-dynatom/},
  [Accessed: 20-May-2020].

\bibitem{canon2019precision}
{Canon {M}edical {S}ystems {USA}, {I}nc.}, ``Aquilon{\texttrademark} precision,
  {ULTRA} {H}igh {R}esolution {CT},''
  \url{https://us.medical.canon/products/computed-tomography/aquilion-precision/},
  [Accessed: 20-May-2020].

\bibitem{natterer2001mathematics}
F.~Natterer, \emph{The mathematics of computerized tomography}.\hskip 1em plus
  0.5em minus 0.4em\relax SIAM, 2001.

\bibitem{feldkamp1984practical}
L.~Feldkamp, L.~Davis, and J.~Kress, ``Practical cone-beam algorithm,''
  \emph{JOSA A}, vol.~1, no.~6, pp. 612--619, 1984.

\bibitem{katsevich2003general}
A.~Katsevich, ``A general scheme for constructing inversion algorithms for cone
  beam {CT},'' \emph{International Journal of Mathematics and Mathematical
  Sciences}, vol. 2003, no.~21, pp. 1305--1321, 2003.

\bibitem{pan2009commercial}
X.~Pan, E.~Y. Sidky, and M.~Vannier, ``Why do commercial {CT} scanners still
  employ traditional, filtered back-projection for image reconstruction?''
  \emph{Inverse problems}, vol.~25, no.~12, p. 123009, 2009.

\bibitem{rudin1992nonlinear}
L.~I. Rudin, S.~Osher, and E.~Fatemi, ``Nonlinear total variation based noise
  removal algorithms,'' \emph{Physica D: Nonlinear Phenomena}, vol.~60, no.
  1-4, pp. 259--268, 1992.

\bibitem{bredies2010total}
K.~Bredies, K.~Kunisch, and T.~Pock, ``Total generalized variation,''
  \emph{SIAM Journal on Imaging Sciences}, vol.~3, no.~3, pp. 492--526, 2010.

\bibitem{sidky2008image}
E.~Y. Sidky and X.~Pan, ``Image reconstruction in circular cone-beam computed
  tomography by constrained, total-variation minimization,'' \emph{Physics in
  medicine and biology}, vol.~53, no.~17, p. 4777, 2008.

\bibitem{jia2010gpu}
X.~Jia, Y.~Lou, R.~Li, W.~Y. Song, and S.~B. Jiang, ``{GPU}-based fast cone
  beam {CT} reconstruction from undersampled and noisy projection data via
  total variation,'' \emph{Medical physics}, vol.~37, no.~4, pp. 1757--1760,
  2010.

\bibitem{0031-9155-59-12-2997}
S.~Niu, Y.~Gao, Z.~Bian, J.~Huang, W.~Chen, G.~Yu, Z.~Liang, and J.~Ma,
  ``Sparse-view {X-ray CT} reconstruction via total generalized variation
  regularization,'' \emph{Physics in Medicine \& Biology}, vol.~59, no.~12, p.
  2997, 2014.

\bibitem{elbakri2003efficient}
I.~A. Elbakri and J.~A. Fessler, ``Efficient and accurate likelihood for
  iterative image reconstruction in {X}-ray computed tomography,'' in
  \emph{Medical Imaging 2003: Image Processing}, vol. 5032.\hskip 1em plus
  0.5em minus 0.4em\relax International Society for Optics and Photonics, 2003,
  pp. 1839--1850.

\bibitem{l2012filtered}
G.~L.~Zeng, ``A filtered backprojection algorithm with characteristics of the
  iterative landweber algorithm,'' \emph{Medical physics}, vol.~39, no.~2, pp.
  603--607, 2012.

\bibitem{nielsen2012filter}
T.~Nielsen, S.~Hitziger, M.~Grass, and A.~Iske, ``Filter calculation for
  {X-ray} tomosynthesis reconstruction,'' \emph{Physics in medicine and
  biology}, vol.~57, no.~12, p. 3915, 2012.

\bibitem{batenburg2012fast}
K.~J. Batenburg and L.~Plantagie, ``Fast approximation of algebraic
  reconstruction methods for tomography,'' \emph{IEEE Transactions on Image
  Processing}, vol.~21, no.~8, pp. 3648--3658, 2012.

\bibitem{pelt2014improving}
D.~M. Pelt and K.~J. Batenburg, ``Improving filtered backprojection
  reconstruction by data-dependent filtering,'' \emph{IEEE Transactions on
  Image Processing}, vol.~23, no.~11, pp. 4750--4762, 2014.

\bibitem{lagerwerf2020automated}
M.~J. Lagerwerf, W.~J. Palenstijn, H.~Kohr, and K.~J. Batenburg, ``Automated
  {FDK-filter selection} for {Cone-beam} {CT} in research environments,''
  \emph{IEEE Transactions on Computational Imaging}, vol. Early acces, 2020.

\bibitem{kunze2007filter}
\BIBentryALTinterwordspacing
H.~Kunze, W.~Haerer, J.~Orman, T.~Mertelmeier, and K.~Stierstorfer, ``Filter
  determination for tomosynthesis aided by iterative reconstruction
  techniques,'' in \emph{9th International Meeting on Fully Three-Dimensional
  Image Reconstruction in Radiology and Nuclear Medicine}, 2007, pp. 309--312.
  [Online]. Available:
  \url{http://www.fully3d.org/2007/Fully3D_HPIR_Proceedings.pdf}
\BIBentrySTDinterwordspacing

\bibitem{jin2017deep}
K.~H. Jin, M.~T. McCann, E.~Froustey, and M.~Unser, ``Deep convolutional neural
  network for inverse problems in imaging,'' \emph{IEEE Transactions on Image
  Processing}, vol.~26, no.~9, pp. 4509--4522, 2017.

\bibitem{pelt2018improving}
D.~M. Pelt, K.~J. Batenburg, and J.~Sethian, ``Improving tomographic
  reconstruction from limited data using mixed-scale dense convolutional neural
  networks,'' \emph{Journal of Imaging}, vol.~4, no.~11, p. 128, 2018.

\bibitem{kida2018cone}
S.~Kida, T.~Nakamoto, M.~Nakano, K.~Nawa, A.~Haga, J.~Kotoku, H.~Yamashita, and
  K.~Nakagawa, ``Cone beam computed tomography image quality improvement using
  a deep convolutional neural network,'' \emph{Cureus}, vol.~10, no.~4, 2018.

\bibitem{pelt2018mixed}
D.~M. Pelt and J.~A. Sethian, ``A mixed-scale dense convolutional neural
  network for image analysis,'' \emph{Proceedings of the National Academy of
  Sciences}, vol. 115, no.~2, pp. 254--259, 2018.

\bibitem{wang2018image}
G.~Wang, J.~C. Ye, K.~Mueller, and J.~A. Fessler, ``Image reconstruction is a
  new frontier of machine learning,'' \emph{IEEE transactions on medical
  imaging}, vol.~37, no.~6, pp. 1289--1296, 2018.

\bibitem{cciccek20163d}
{\"O}.~{\c{C}}i{\c{c}}ek, A.~Abdulkadir, S.~S. Lienkamp, T.~Brox, and
  O.~Ronneberger, ``{3D U-Net}: learning dense volumetric segmentation from
  sparse annotation,'' in \emph{International conference on medical image
  computing and computer-assisted intervention}.\hskip 1em plus 0.5em minus
  0.4em\relax Springer, 2016, pp. 424--432.

\bibitem{ronneberger2015u}
O.~Ronneberger, P.~Fischer, and T.~Brox, ``U-net: {Convolutional} networks for
  biomedical image segmentation,'' in \emph{International Conference on Medical
  image computing and computer-assisted intervention}.\hskip 1em plus 0.5em
  minus 0.4em\relax Springer, 2015, pp. 234--241.

\bibitem{bishop2006pattern}
C.~M. Bishop, \emph{Pattern recognition and machine learning}.\hskip 1em plus
  0.5em minus 0.4em\relax Springer Science+ Business Media, 2006.

\bibitem{pelt2013fast}
D.~M. Pelt and K.~J. Batenburg, ``Fast tomographic reconstruction from limited
  data using artificial neural networks,'' \emph{IEEE Transactions on Image
  Processing}, vol.~22, no.~12, pp. 5238--5251, 2013.

\bibitem{van1990sirt}
A.~Van~der Sluis and H.~A. van~der Vorst, ``{SIRT-and CG-type} methods for the
  iterative solution of sparse linear least-squares problems,'' \emph{Linear
  Algebra and its Applications}, vol. 130, pp. 257--303, 1990.

\bibitem{kang-2017-deep-convol}
\BIBentryALTinterwordspacing
E.~Kang, J.~Min, and J.~C. Ye, ``A deep convolutional neural network using
  directional wavelets for low-dose {X}-ray {CT} reconstruction,''
  \emph{Medical Physics}, vol.~44, no.~10, pp. e360--e375, Oct 2017. [Online].
  Available: \url{https://doi.org/10.1002/mp.12344}
\BIBentrySTDinterwordspacing

\bibitem{adler2017solving}
J.~Adler and O.~{\"O}ktem, ``Solving ill-posed inverse problems using iterative
  deep neural networks,'' \emph{Inverse Problems}, vol.~33, no.~12, p. 124007,
  2017.

\bibitem{adler2018learned}
------, ``Learned primal-dual reconstruction,'' \emph{IEEE transactions on
  medical imaging}, vol.~37, no.~6, pp. 1322--1332, 2018.

\bibitem{kobler2017variational}
E.~Kobler, T.~Klatzer, K.~Hammernik, and T.~Pock, ``Variational networks:
  connecting variational methods and deep learning,'' in \emph{German
  conference on pattern recognition}.\hskip 1em plus 0.5em minus 0.4em\relax
  Springer, 2017, pp. 281--293.

\bibitem{hammernik2018learning}
K.~Hammernik, T.~Klatzer, E.~Kobler, M.~P. Recht, D.~K. Sodickson, T.~Pock, and
  F.~Knoll, ``Learning a variational network for reconstruction of accelerated
  mri data,'' \emph{Magnetic resonance in medicine}, vol.~79, no.~6, pp.
  3055--3071, 2018.

\bibitem{venkatakrishnan-2013-plug-and}
\BIBentryALTinterwordspacing
S.~V. Venkatakrishnan, C.~A. Bouman, and B.~Wohlberg, ``Plug-and-play priors
  for model based reconstruction,'' \emph{2013 IEEE Global Conference on Signal
  and Information Processing}, Dec 2013. [Online]. Available:
  \url{https://doi.org/10.1109/globalsip.2013.6737048}
\BIBentrySTDinterwordspacing

\bibitem{romano-2016-littl-engin}
\BIBentryALTinterwordspacing
Y.~Romano, M.~Elad, and P.~Milanfar, ``The little engine that could:
  Regularization by denoising ({RED}),'' \emph{SIAM Journal on Imaging
  Sciences}, vol.~10, no.~4, pp. 1804--1844, Jan 2017. [Online]. Available:
  \url{https://doi.org/10.1137/16m1102884}
\BIBentrySTDinterwordspacing

\bibitem{reehorst-2018-regul-by-denois}
\BIBentryALTinterwordspacing
E.~T. Reehorst and P.~Schniter, ``Regularization by denoising: Clarifications
  and new interpretations,'' \emph{CoRR}, 2018. [Online]. Available:
  \url{http://arxiv.org/abs/1806.02296v1}
\BIBentrySTDinterwordspacing

\bibitem{lunz2018adversarial}
S.~Lunz, O.~{\"O}ktem, and C.-B. Sch{\"o}nlieb, ``Adversarial regularizers in
  inverse problems,'' in \emph{Advances in Neural Information Processing
  Systems}, 2018, pp. 8507--8516.

\bibitem{mukherjee2020learned}
S.~Mukherjee, S.~Dittmer, Z.~Shumaylov, S.~Lunz, O.~{\"O}ktem, and C.-B.
  Sch{\"o}nlieb, ``Learned convex regularizers for inverse problems,''
  \emph{arXiv preprint arXiv:2008.02839}, 2020.

\bibitem{shelhamer-2017-fully-convol}
\BIBentryALTinterwordspacing
E.~Shelhamer, J.~Long, and T.~Darrell, ``Fully convolutional networks for
  semantic segmentation,'' \emph{IEEE Transactions on Pattern Analysis and
  Machine Intelligence}, vol.~39, no.~4, pp. 640--651, 2017. [Online].
  Available: \url{https://doi.org/10.1109/tpami.2016.2572683}
\BIBentrySTDinterwordspacing

\bibitem{perone-2018-spinal-cord}
\BIBentryALTinterwordspacing
C.~S. Perone, E.~Calabrese, and J.~Cohen-Adad, ``Spinal cord gray matter
  segmentation using deep dilated convolutions,'' \emph{Scientific Reports},
  vol.~8, no.~1, Apr 2018. [Online]. Available:
  \url{https://doi.org/10.1038/s41598-018-24304-3}
\BIBentrySTDinterwordspacing

\bibitem{zhang-2017-beyon-gauss-denois}
\BIBentryALTinterwordspacing
K.~Zhang, W.~Zuo, Y.~Chen, D.~Meng, and L.~Zhang, ``Beyond a gaussian denoiser:
  Residual learning of deep {CNN} for image denoising,'' \emph{IEEE
  Transactions on Image Processing}, vol.~26, no.~7, pp. 3142--3155, Jul 2017.
  [Online]. Available: \url{https://doi.org/10.1109/tip.2017.2662206}
\BIBentrySTDinterwordspacing

\bibitem{ye2018deep}
J.~C. Ye, Y.~Han, and E.~Cha, ``Deep convolutional framelets: A general deep
  learning framework for inverse problems,'' \emph{SIAM Journal on Imaging
  Sciences}, vol.~11, no.~2, pp. 991--1048, 2018.

\bibitem{anthony2009neural}
M.~Anthony and P.~L. Bartlett, \emph{Neural network learning: {Theoretical}
  foundations}.\hskip 1em plus 0.5em minus 0.4em\relax cambridge university
  press, 2009.

\bibitem{levenberg1944method}
K.~Levenberg, ``A method for the solution of certain non-linear problems in
  least squares,'' \emph{Quarterly of applied mathematics}, vol.~2, no.~2, pp.
  164--168, 1944.

\bibitem{marquardt1963algorithm}
D.~W. Marquardt, ``An algorithm for least-squares estimation of nonlinear
  parameters,'' \emph{Journal of the society for Industrial and Applied
  Mathematics}, vol.~11, no.~2, pp. 431--441, 1963.

\bibitem{kingma2014adam}
D.~P. Kingma and J.~Ba, ``Adam: A method for stochastic optimization,''
  \emph{arXiv preprint arXiv:1412.6980}, 2014.

\bibitem{kudo1998cone}
H.~Kudo, F.~Noo, and M.~Defrise, ``Cone-beam filtered-backprojection algorithm
  for truncated helical data,'' \emph{Physics in Medicine \& Biology}, vol.~43,
  no.~10, p. 2885, 1998.

\bibitem{hubbell1995tables}
J.~H. Hubbell and S.~M. Seltzer, ``Tables of {X-ray} mass attenuation
  coefficients and mass energy-absorption coefficients 1 {keV} to 20 {MeV} for
  elements {Z= 1} to 92 and 48 additional substances of dosimetric interest,''
  National Inst. of Standards and Technology-PL, Gaithersburg, MD (United
  States). Ionizing Radiation Div., Tech. Rep., 1995.

\bibitem{coban2020explorative}
S.~B. Coban, F.~Lucka, W.~J. Palenstijn, D.~Van~Loo, and K.~J. Batenburg,
  ``Explorative imaging and its implementation at the {FleX-ray} laboratory,''
  \emph{Journal of Imaging}, vol.~6, no.~4, p.~18, 2020.

\bibitem{walnuts}
\BIBentryALTinterwordspacing
M.~J. Lagerwerf, S.~B. Coban, and K.~J. Batenburg, ``{High-resolution cone-beam
  scan of twenty-one walnuts with two dosage levels},'' Apr. 2020. [Online].
  Available: \url{https://doi.org/10.5281/zenodo.3763412}
\BIBentrySTDinterwordspacing

\bibitem{wang2004image}
Z.~Wang, A.~C. Bovik, H.~R. Sheikh, and E.~P. Simoncelli, ``Image quality
  assessment: from error visibility to structural similarity,'' \emph{IEEE
  transactions on image processing}, vol.~13, no.~4, pp. 600--612, 2004.

\bibitem{scikit-image}
S.~van~der Walt, J.~L. {S}ch\"onberger, J.~{Nunez-Iglesias}, F.~{B}oulogne,
  J.~D. {W}arner, N.~{Y}ager, E.~{G}ouillart, T.~{Y}u, and the scikit-image
  contributors, ``scikit-image: image processing in {P}ython,'' \emph{PeerJ},
  vol.~2, p. e453, 6 2014.

\bibitem{bernard20203d}
A.~Bernard, S.~Hamdy, L.~Le~Corre, E.~Dirlewanger, and F.~Lheureux, ``{3D}
  characterization of walnut morphological traits using {X}-ray computed
  tomography,'' \emph{preprint}, 2020.

\bibitem{dice1945measures}
L.~R. Dice, ``Measures of the amount of ecologic association between species,''
  \emph{Ecology}, vol.~26, no.~3, pp. 297--302, 1945.

\bibitem{walt2011numpy}
S.~v.~d. Walt, S.~C. Colbert, and G.~Varoquaux, ``The {NumPy} array: a
  structure for efficient numerical computation,'' \emph{Computing in Science
  \& Engineering}, vol.~13, no.~2, pp. 22--30, 2011.

\bibitem{odl}
J.~Adler, H.~Kohr, and O.~Öktem, ``Odl 0.6.0,'' Apr. 2017.

\bibitem{van2016fast}
W.~van Aarle, W.~J. Palenstijn, J.~Cant, E.~Janssens, F.~Bleichrodt,
  A.~Dabravolski, J.~De~Beenhouwer, K.~J. Batenburg, and J.~Sijbers, ``Fast and
  flexible {X-ray} tomography using the {ASTRA} toolbox,'' \emph{Optics
  express}, vol.~24, no.~22, pp. 25\,129--25\,147, 2016.

\bibitem{frigo2005design}
M.~Frigo and S.~G. Johnson, ``The design and implementation of {FFTW3},''
  \emph{Proceedings of the IEEE}, vol.~93, no.~2, pp. 216--231, 2005.

\bibitem{behnel2010cython}
S.~Behnel, R.~Bradshaw, C.~Citro, L.~Dalcin, D.~Seljebotn, and K.~Smith,
  ``Cython: {The Best of Both Worlds},'' \emph{Computing in Science
  Engineering}, vol.~13, no.~2, pp. 31--39, 2011.

\bibitem{NEURIPS2019_9015}
\BIBentryALTinterwordspacing
A.~Paszke, S.~Gross, F.~Massa, A.~Lerer, J.~Bradbury, G.~Chanan, T.~Killeen,
  Z.~Lin, N.~Gimelshein, L.~Antiga, A.~Desmaison, A.~Kopf, E.~Yang, Z.~DeVito,
  M.~Raison, A.~Tejani, S.~Chilamkurthy, B.~Steiner, L.~Fang, J.~Bai, and
  S.~Chintala, ``{PyTorch}: An imperative style, high-performance deep learning
  library,'' in \emph{Advances in Neural Information Processing Systems 32},
  H.~Wallach, H.~Larochelle, A.~Beygelzimer, F.~d'Alch\'{e} Buc, E.~Fox, and
  R.~Garnett, Eds.\hskip 1em plus 0.5em minus 0.4em\relax Curran Associates,
  Inc., 2019, pp. 8024--8035. [Online]. Available:
  \url{http://papers.neurips.cc/paper/9015-pytorch-an-imperative-style-high-performance-deep-learning-library.pdf}
\BIBentrySTDinterwordspacing

\bibitem{hendriksen2019fly}
A.~A. Hendriksen, D.~M. Pelt, W.~J. Palenstijn, S.~B. Coban, and K.~J.
  Batenburg, ``On-the-fly machine learning for improving image resolution in
  tomography,'' \emph{Applied Sciences}, vol.~9, no.~12, p. 2445, 2019.

\bibitem{nnfdk}
M.~J. Lagerwerf, ``Neural network {FDK} algorithm,''
  \url{https://github.com/MJLagerwerf/nn_fdk}, [Accessed: 20-May-2020].

\bibitem{nguyen1990truck}
D.~Nguyen and B.~Widrow, ``The truck backer-upper: {An} example of
  self-learning in neural networks,'' in \emph{Advanced neural
  computers}.\hskip 1em plus 0.5em minus 0.4em\relax Elsevier, 1990, pp.
  11--19.

\end{thebibliography}
\end{document}